\documentclass[prb,twocolumn,showpacs,amsmath,amssymb]{revtex4}
\usepackage{graphicx}% Include figure files
\usepackage{dcolumn}% Align table columns on decimal point
\usepackage{bm}% bold math

\begin{document}

\preprint{APS/123-QED}

\title{Angular dependent planar metamagnetism in \\the hexagonal compounds TbPtIn and TmAgGe}% Force line breaks with \\

\author{E. Morosan, S. L. Bud'ko and P. C. Canfield}
\affiliation{Ames Laboratory and Department of Physics and
Astronomy, \\Iowa State University, Ames, IA 50011, USA}

\date{\today}% It is always \today, today,

\begin{abstract}
Detailed magnetization measurements, M(T,H,$\theta$), were
performed on single crystals of TbPtIn and TmAgGe (both members of
the hexagonal Fe$_2$P/ZrNiAl structure type), for the magnetic
field H applied perpendicular to the crystallographic c axis.
These data allowed us to identify, for each compound, the
easy-axes for the magnetization, which coincided with high
symmetry directions ([120] for TbPtIn and [110] for TmAgGe). For
fixed orientations of the field along each of the two six-fold
symmetry axes, a number of magnetically ordered phases is being
revealed by M(H,T) measurements below $T_N$. Moreover,
$T~\simeq~2$ K,  M(H)$|_{\theta}$ measurements for both compounds
(with H applied parallel to the basal plane), as well as $T~=~20$
K data for TbPtIn, reveal five metamagnetic transitions with
simple angular dependencies: $H_{c
i,j}~\sim~1/\cos(\theta~\pm~\varphi)$, where $\varphi~=~0^0$ or
$60^0$. The high field magnetization state varies with $\theta$
like $2/3*\mu_{sat}($R$^{3+})*\cos\theta$, and corresponds to a
crystal field limited saturated paramagnetic, CL-SPM, state.
Analysis of these data allowed us to model the angular dependence
of the locally saturated magnetizations $M_{sat}$ and critical
fields $H_c$ with a three coplanar Ising-like model, in which the
magnetic moments are assumed to be parallel to three adjacent easy
axes. Furthermore, net distributions of moments were inferred
based on the measured data and the proposed model.

\end{abstract}

\pacs{75.25.+z; 75.10.-b; 75.30.Gw; 75.30.Kz; 75.50.Ee}

\maketitle

\section{\label{sec:intro}Introduction}% force lowercase using \lowercase{text}

Numerous studies of angular dependent local moment metamagnetism
show that simple geometric relationships exist between the
critical fields of the metamagnetic phase transitions, and the
angle that the applied field makes with the corresponding easy
axis. One simple case is that of TbNi$_{2}$Ge$_{2}$\cite{bud01}, a
tetragonal compound with Tb ions in tetragonal point symmetry,
where, at low temperatures, the crystalline electric field (CEF)
anisotropy confines the local moments along the c ([001])
crystallographic axis (Ising-like system). Consequently, several
metamagnetic transitions are observed, with the critical field
values $H_{c}$ having a $1/\cos\theta$ dependence on the angle
between the applied field \textit{H} and the c-axis. A more
complex situation is encountered when the magnetic moments are
allowed more degrees of freedom , \textit{i.e.} when the CEF
anisotropy constrains them to an easy plane. This is the case in
the tetragonal compounds RNi$_2$B$_2$C\cite{choTb,can02,can03},
with R = Tb - Er, and RAgSb$_{2}$\cite{mye03} for R = Dy. The R
ions are again in tetragonal point symmetry and the local moments
are confined to four equivalent [110] or [100] crystallographic
directions; thus the angular dependent magnetization measurements,
when the field is applied in the basal plane, reveal the four-fold
anisotropy of the longitudinal magnetization that reflects the
symmetry of the unit cell. The angular dependencies of the locally
saturated magnetizations $M_{sat}$ and the critical fields $H_{c}$
could be treated by simple analysis, and plausible net
distribution of moments could be inferred for each metamagnetic
phase\cite{can02,mye03}. Kalatsky and Pokrovsky\cite{kal04}
elaborated the four-position clock model, which agrees well with
the observed metamagnetism in HoNi$_2$B$_2$C and DyAgSb$_{2}$.

Initially thought to be improbable, angular dependent
metamagnetism in extremely planar systems is now an accepted and
understandable event in tetragonal compounds; but this has not yet
been well studied in hexagonal compounds, and this is the
motivation for the present analysis. Recently we reported the
presence of metamagnetism in a hexagonal class of materials,
\textit{i.e.} the RAgGe compounds, for R = Tb - Tm\cite{mor05}.
They crystallize in the ZrNiAl-type structure, an ordered variant
of the hexagonal Fe$_{2}$P family. In this structure, there is a
unique rare earth site in the unit cell, with the rare earth ions
occupying equivalent 3g positions with orthorhombic point symmetry
(fig.1). The anisotropic susceptibility was found to be axial for
TbAgGe and progressed towards extremely planar for TmAgGe. The
physical properties of the TmAgGe compound (antiferromagnetic
ordering in the ground state, extremely planar anisotropy,
metamagnetism when the applied field was perpendicular to the
c-axis) are very similar to those of the isostructural TbPtIn
system, previously only known in polycrystalline form\cite{gal10,
wat11}, making both compounds good candidates for a study of the
angular dependent metamagnetism in hexagonal systems. In both
TmAgGe and TbPtIn the R ions occupy the same site, whereas both
ligands are different. Having two such systems will allow us to
show that the behavior we find is not specific to one compound,
but a more general result associated with this structure, or
perhaps with the orthorhombic point symmetry in a hexagonal unit
cell. As we shall see, the N\'{e}el temperature is much higher for
TbPtIn (46.0 K) than for TmAgGe (4.2 K), with the former also
showing a possible spin reorientation transition at a fairly high
temperature ($T_m~=~27.4$ K). This latter transition in TbPtIn was
missed by the measurements on polycrystalline samples\cite{wat11},
where even the nature of the magnetically ordered state below
$\sim~50$ K was not identified. Thus we can once again emphasize
the advantage of analysis on single crystals rather than on
polycrystalline samples.

This paper is organized as follows: after a brief description of
the experimental methods, we present the \textit{M(T,H)} data on
TbPtIn, emphasizing the complexity of its ordered state, for field
orientations along the two in-plane, high symmetry directions;
this is followed by the M(H,$\theta$) measurements at low
temperature, from which the values of the locally saturated
magnetizations and critical fields as a function of $\theta$ were
extracted. Next, we will introduce a model for the net
distribution of moments, which we subsequently use to calculate
the expected, locally saturated magnetizations $M_{j}$ and the
critical fields (for the transition from state i to state j)
$H_{ci,j}$ as functions of $\theta$. This will be followed by a
comparison of how the calculated and experimental $M_{j}$ and
$H_{ci,j}$ values vary with the angle between the applied field H
and the easy axis for this compound.

Similar measurements performed on TmAgGe will then be shown,
leading to the corresponding experimental $M_{j}$($\theta$) and
$H_{ci,j}$($\theta$) phase diagrams, which we will again map onto
the appropriate model calculation.

We will also analyze similar $M(H,\theta)$ data of TbPtIn for
$T~=~20$ K, and use the same model in order to characterize a
magnetic phase present only at higher temperatures.

Finally, we will summarize our key results and also indicate how
our model can be generalized to a variety of possible point
symmetries in tetragonal and hexagonal systems.

\section{Experimental methods}

Using an initial concentration of
Tb$_{0.05}$Pt$_{0.05}$In$_{0.90}$, clean hexagonal rods of TbPtIn
were grown; this In-rich self-flux was used because of its
low-melting temperature and because it introduces no new elements
into the melt. The constituent elements were placed in alumina
crucibles and sealed in quartz tubes under partial argon pressure;
subsequently they were heated up to $1200^{0}$ C, and then cooled
down to $800^{0}$ C over 48 h. Removal of the excess liquid
solution revealed hexagonal rods with the c-axis along the axis of
the rod. The crystal structure was confirmed by room temperature
powder x-ray diffraction measurements, using Cu K$_{\alpha}$
radiation, from which the lattice parameters a
$=~(7.55~\pm~0.01)~\AA$ and c $=~(3.88~\pm~0.01)~\AA$ were also
obtained. It is worth noting that the above temperature profile
yielded in most cases well-formed, fairly good crystals (residual
resistivity ratio RRR $=~5.2$); however, some of the crystals had
hollow channels in the center, sometimes with flux inclosures, and
the quality of the resulting hexagonal rods seems to only
deteriorate with slower cooling rates (\textit{i.e.} $400^{0}$ C /
100h). Single crystals of TmAgGe were also flux-grown out of Ag-Ge
self flux, as described in detail in Morosan \textit{et
al.}\cite{mor05}.

Magnetic measurements were performed in a Quantum Design Magnetic
Properties Measurement System (MPMS) SQUID magnetometer, with a
specially modified sample holder to allow the rotation of the
sample so that its c-axis stays perpendicular to the applied
magnetic field. Additional measurements up to 140 kG were also
taken, using an extraction magnetometer in a Quantum Design
Physical Properties Measurement System (PPMS). In order to avoid
torque on the rotator due to the extreme anisotropy of the samples
, ~small mass samples ~(~\textit{i.e.}  ~m $=~(0.40~\pm~0.05)$ mg
in the case of ~TbPtIn,~~~~and~~~m $=~(0.54~\pm~0.05)$ mg for
TmAgGe) were used for the angular dependent measurements. To
correct for the large weighing errors that would result from such
small masses, we used a 13.55 mg TbPtIn sample, and a 2.40 mg
TmAgGe sample respectively, to measure \textit{M(H)} curves for
$\theta~=~0^{0}$ and $30^{0}$ (with the angle $\theta$ measured
between the applied field H and the easy axis, as described
below). Whereas some errors are still introduced by this
calibration due to the manual orientation of the large samples,
the angular accuracy with which we are able to orient them is
probably within $10^0$. The various \textit{M(H}, $\theta$)
measurements on the smaller samples were than calibrated using the
data for the larger mass pieces; this is believed to be very
accurate given that the angular uncertainty in the rotator is less
than $1^{0}$. There is additional uncertainty introduced by
possible misalignment of the small piece with the c-axis exactly
perpendicular to the applied field (due to the construction of the
sample holder, this misalignment shouldn't be more than $10^{0}$).

Laue experiments have been performed on well-formed single
crystals. They confirmed that the as-grown facettes of the
hexagonal rods are alternatingly parallel to the (101) and (011)
crystallographic planes (the macroscopic facettes are aligned with
the hexagonal unit cell axes [100] or [010]).

\section{T\lowercase{b}P\lowercase{t}I\lowercase{n}}

Fig.2 shows the inverse magnetic susceptibility for TbPtIn, as
well as the low-temperature magnetization in the inset. This
compound appears to order antiferromagnetically below
$T_{N}~=~46.0$ K, with what is probably a spin-reorientation or a
commensurate-to-incommensurate transition around $T_m~=~27.4$ K,
as indicated by the peaks in the $d(M*T/H)/dT$ plot\cite{chiT} in
Fig.3a. In the determination of these temperature values, $M/H$
represents the polycrystalline average susceptibility
$\chi_{ave}$, calculated as

$\chi_{ave}~=~\frac{1}{3}~\chi_{c}+\frac{2}{3}~\chi_{ab}$

or, when measurements have been done along all three salient
directions,

$\chi_{ave}~=~\frac{1}{3}~(\chi_{[001]}+\chi_{[100]}+\chi_{[120]})$.

The above temperatures are further confirmed by the $C_{p}(T)$ and
$\rho(T)$ ($i||ab$) data, also shown in Fig.3b and c.

In the ordered state, as well as in the paramagnetic state up to
$\sim100$ K, the susceptibility is extremely anisotropic, with the
local Tb moments confined to the \textit{ab}-plane at low
temperatures. To check the origin of the anisotropy in the
paramagnetic state, single crystals of YPtIn were grown, with a
small number of non-magnetic Y$^{3+}$ ions substituted with
magnetic Tb$^{3+}$ ions. The extremely anisotropic susceptibility
and magnetization of the diluted compound demonstrate that this is
single-ion anisotropy associated with the CEF splitting of the
Hund's rule ground state J multiplet. From the Curie-Weiss
effective moment, as determined from the average inverse
susceptibility (Fig.4), the concentration x of the diluted
compound (Tb$_{x}$Y$_{1-x})$PtIn is $x~=~0.024$, whereas the high
field magnetization equals $\sim~5.6~\mu_B/$Tb$^{3+}$. (If we
assume the high field magnetization to be $\sim~6.25
\mu_B/$Tb$^{3+}$, as expected based on the model described below,
the resulting concentration will be $x~=~0.019$).

For temperatures higher than 150 K, Curie-Weiss behavior of the
pure TbPtIn compound can be inferred (Fig.2) from the linear
inverse susceptibilities, resulting in anisotropic Weiss
temperatures $\Theta_{ab}~=~39.0$ K and $\Theta_{c}~=~23.3$ K. The
polycrystalline average susceptibility $\chi_{ave}$ yielded an
effective moment $\mu_{eff}~=~9.74 \mu_{B}$/Tb$^{3+}$, very close
to the theoretical value $9.72~\mu_{B}$; the corresponding Weiss
temperature is $\Theta_{ave}~=~33.3$K.

The field dependent magnetization measurements shown in Fig.5a not
only confirm the in-plane/out-of-plane anisotropy observed in the
ordered state, but also indicate anisotropic magnetization within
the basal plane. Moreover, several metamagnetic transitions can be
seen for the field parallel to the \textit{ab}-plane, for fields
up to 140 kG in the $M(H)$ data, and for the magnetoresistance
measurements up to 90 kG in Fig.5b. The geometry of the crystals
led to more uncertainty in orienting the resistance pieces than
the ones for magnetization measurements; therefore we can infer
the approximate orientation of the magnetoresistance sample with
respect to the applied field, by comparing the
$\Delta\rho(H)/\rho(0)$ data (Fig.5b) with the $M(H)$ curves for
$H||ab$ (Fig.5a): since various features in the magnetoresistance
measurements occur closer to the critical fields in $M([120])$, we
can assume that the field was almost parallel to the [120]
direction. (The sharp drop in $\rho(H)$ below $\sim1$ kG (Fig.5b)
is very likely due to superconductivity of residual In flux on the
surface of our resistance bar). For \textit{H} along the c-axis,
the magnetization increases almost linearly with increasing field
(Fig.5a), while staying far smaller than $M_{ab}$.

At the highest applied field ($H~=~140$ kG), the magnetization
values for the three shown orientations are $M([110])~=~5.86
\mu_{B}$/Tb, $M([120])~=~6.45 \mu_{B}$/Tb and $M([001])~=~0.92
\mu_{B}$/Tb. Whereas the extreme planar anisotropy of TbPtIn and
the anisotropy within the \textit{ab}-plane recommended this
compound for a study of the angular dependent metamagnetism, the
fact that the magnetization values were smaller in all three
directions than the calculated $\mu_{sat}~=~9 \mu_{B}$ for
$Tb^{3+}$ ions is somewhat intriguing. One plausible explanation
for the low magnetization values would be the existence of more
metamagnetic transitions for fields unaccessible with our
measurement systems (\textit{i.e.} above 140 kG). Another
possibility is that an additional energy scale (such as CEF
splitting) exists, that confines the three local moments to three
distinct, non-collinear, in-plane orientations. As shall be shown
below, we believe the latter to be the more likely scenario.

In order to determine the easy axes of the system, we continuously
rotated a small piece of the diluted sample
(Tb$_{x}$Y$_{1-x})$PtIn ($x \approx 0.02$) in an applied field
$H~=~55$ kG (perpendicular to \textit{c}); for constant
temperature $T~=~2$ K, the corresponding magnetization
measurement, shown as open symbols in Fig.6, roughly follows a
$\cos\theta$ dependence (solid line) around the closest
[120]-equivalent directions (\textit{i.e.} $\theta~=~60^0*n$,
$n~=~integer$), where the maxima occur. As we have seen before,
the concentration of the diluted sample is
$x~=~(0.0215~\pm~0.0025)$; given this uncertainty, the absolute
value of the magnetization $M(\theta)$ in Fig.6 could not be
determined, and thus we scaled the data to the maximum value,
$M(\theta~=~0^0)$. Similar behavior appears in the pure TbPtIn
compound (full symbols in Fig.6), where the measured data have
also been scaled to their corresponding maximum value at
$\theta~=~0^0$. We can conclude that the easy axes of the TbPtIn
system are the [120] directions. However dramatic departures from
the $\cos\theta$ angular dependence can be noticed. The
magnetization for TbPtIn, indicative of strong interactions
between the local moments, is also consistent with various
metamagnetic states crossing the $H~=~55$ kG line at different
angles. Based on the above data, we will consider the easy axes to
be the [120]-equivalent directions, and the angle $\theta$ will be
measured from the closest easy axis.

In order to get an idea about the various metamagnetic states in
this compound, we first explored changes of the critical fields
and temperatures for two fixed orientations. The corresponding
$M(T)\mid_{H,\theta}$ and $M(H)\mid_{T,\theta}$ measurements shown
in Fig.7-8 have been used to determine the $H-T$ phase diagrams
for the two in-plane high symmetry directions, $H||[120]$ and
$H||[110]$ respectively. As illustrated in the insets in these
figures, the points in these phase diagrams have been determined
from local maxima in $d(M*T)/dT$ (full circles in fig.9) for fixed
fields, and in $dM/dH$ (open circles in fig.9) for various $M(H)$
isotherms. Even though in Fisher \textit{et al.}\cite{chiT} the
maxima in $d(M*T)/dT$ criterion is described only for
antiferromagnetic systems, we apply it here not only for the AF
state, but also for high-field states, where the magnetization has
a net ferromagnetic component. We are confident that small errors
are thus introduced, given the consistency of the critical field
and temperature values obtained from both $d(M*T)/dT$ and $dM/dH$
derivatives (full and open symbols respectively in Fig.9). Given
that the transition peaks were broad for some field and
temperature values, we used Lorentzian fits of the corresponding
derivatives (thick lines in fig.7 and 8, insets) to determine the
critical values $H_c$ and $T_c$.

The resulting $H-T$ phase diagrams for $H||[120]$ (fig.9a) and
$H||[110]$ (fig.9c) are qualitatively similar, at low temperatures
and low fields showing the metamagnetic states already seen in the
\textit{M(H)} data in Fig.5. For $H||[120]$, fig.9a shows that the
antiferromagnetic ground state persists up to about 20 kG,
followed by a small intermediate state $M_{1}$ (between $\sim~20$
kG and 28 kG) and a higher field state $M_{2}$ up to $\sim~54$ kG;
as field is being further increased, the paramagnetic $PM$ state
is reached, as already indicated in fig.5a by the horizontal
plateaux measured up to 140 kG. (At low temperatures, this is a
crystal-field limited saturated paramagnetic CL-SPM state, in
which, as discussed below, all moments are assumed to be in their
'up' positions, while still confined by the strong CEF energy to
three distinct, non-collinear directions within the basal plane.)
When moving up in temperature at low fields, we find the
antiferromagnetic ground state to extend up to $\sim~27.4$ K,
whereas the magnetic ordered state persists up to $\sim~46.0$ K;
both transition temperatures have been already observed (fig.2 and
3). The $M_{1}$ phase exists below $\sim~5.0$ K, after which, for
a limited temperature range (5.0 K $<~T~<~15.0$ K), there is a
direct transition from the $AF$ to the $M_{2}$ state. Between
$15.0$ K and $27.4$ K, or $18.4$ kG and $2.0$ kG respectively,
another intermediate phase, $M_{4}$, forms. The inset in fig.8a
represents an example of two isothermal cuts of the $[120]$ phase
diagram in fig.9a: the phase boundaries of the bubble-like phase
$M_{4}$ are very close in field at constant T, and so the lower
peak in the $T~=~20.0$ K isothermal derivative $dM/dH$ is poorly
defined; the two higher peaks are fairly sharp, similar to the one
defining the $M_{2}$ to $PM$ transition in the $T~=~30.0$ K
isotherm. However, as we move down in field along the latter
isotherm, one broad peak around $15.4$ K may indicate the crossing
of another almost horizontal phase boundary (leading into the
$M_5$ state), and thus hard to identify in $d(M*T)/dT$. One more
peak at $H_c~\simeq~3.4$ kG is indicative of possibly another
phase $AF'$ existing below this field, between $28$ K and $46.0$
K. This is consistent with an antiferromagnetic ordered state
below $46.0$ K, with an incommensurate-commensurate transition
around $28$ K which frequently occurs in intermetallic compounds.

For $H||[110]$ (Fig.9c) the $H-T$ phase diagram is fairly similar,
with only a few differences: a far less distinct $M_{1}$ phase and
a lower upper-boundary for the $M_2$ region. However, the most
notable difference is a new high field phase, $M_3$, whose upper
boundary is determined by the points indicated with small arrows
in fig.7b. (As the field is being increased towards 140 kG, this
line becomes almost vertical, making it difficult to identify also
in $M(H)|_T$ measurements.)

In order to see how the $H-T$ phase diagram evolves from
$H||[120]$ to $H||[110]$, we collected comparable $M(T)|_H$ and
$M(H)|_T$ data for an intermediate orientation of the applied
field (approximately $12^0$ from the easy axis $[120]$). Fig.10a
shows these $M(T)$ curves for $H~=~1-70$ kG, with the small arrows
indicating the highest-T transition at each field value, as
determined from the $d(M*T)/dT$ maxima. Thus, the upper-most phase
boundary in Fig.9b, representing the phase diagram for this
intermediate position, can be followed in field up to $H~=~65$ kG
(full symbols). For this orientation, we can also identify this
line in the $M(H)$ data, and an example is shown in Fig.10b for
the $T~=~10$ K $M(H)$ isotherm and its $dM/dH$ derivative.
Overall, the features common to both Fig.9a and c are also present
in Fig.9b; moreover, in going from the $[110]$ to the $[120]$
direction, the $M_3$ phase is being compressed, such that in the
intermediate position the phase boundary separating it from $PM$
is fully delineated below 70 kG.

Given the clear in-plane anisotropy of the magnetization
(fig.5-10), it becomes desirable to systematically determine the
angular dependence of the critical fields $H_{c i,j}(\theta)$ and
locally saturated magnetizations $M_{\textit{j}}(\theta)$ of the
$H\bot{c}$ metamagnetic transitions at $T~=~2$ K.

Fig.11 shows a series of magnetization isotherms ($T~=~2$ K)
measured at various angles relative to the easy axis. The critical
fields $H_{ci,j}$, for the transition between states i and j, were
determined from maxima in $dM/dH$, as exemplified in fig.12, and
are shown as full symbols in Fig.13a. In most cases, Lorentzian
fits of the derivative peaks were used (solid line in Fig.12) to
more accurately determine the critical field values. The open
symbols in fig.13 represent reflections of the measured data
across the $\theta~=~0^0$ direction; whenever the measured points
extend beyond the $0^0..30^0$ region, they almost coincide with
the calculated reflections, as expected for a symmetry direction.

For $\theta~\leq~12^{0}$ the antiferromagnetic AF ground state
exists for fields up to about 20 kG, after which two closely
spaced metamagnetic transitions occur, with critical fields, at
$\theta~=~0^{0}$, $H_{cAF,1}~=~20.5$ kG (for the AF to $M_1$
transition) and $H_{c1,2}~=~27.7$ kG (corresponding to the
transition from the $M_1$ to the $M_2$ state). A third transition
from $M_2$ to $CL-SPM$, around a critical field
$H_{c2,CL-SPM}~=~53.7$ kG, changes very little with angle up to
$\theta \approx~8^{0}$; for higher angles, another metamagnetic
state $M_{3}$ forms, being delineated by two distinct critical
fields, $H_{c2,3}$ and $H_{c3,CL-SPM}$. As the former decreases
with the angle, the latter soon reaches values around the maximum
field of 55 kG available in the SQUID magnetometer used for these
measurements. In order to follow this latest transition in higher
magnetic fields, additional measurements were taken in a different
magnetometer, for fields up to 70 kG, and a slightly different
temperature ($T~=~1.85$ K); these data are shown in the inset in
Fig.11a, but by $\theta~=~12^{0}$, $H_{c3,CL-SPM}$ becomes larger
than 70 kG, therefore we can only anticipate that this transition
still exists for larger angles. (The $H-T$ phase diagrams (Fig.9b
and c) seem to indicate that this critical field value increases
from $\sim~68$ kG for $\theta \cong~12^{0}$, to more than 140 kG
at $\theta~=~30^{0}$). Slight differences can be noticed between
the data sets taken in the two machines, very likely due to the
different temperatures at which they were taken. A linear scaling
of the two data sets by a factor of $\sim~1.07$ was necessary, for
both the magnetization values and the critical fields; the scaling
of the magnetization values can be explained by an assumption of
slightly different angles between the applied field and the
rotator axis in each magnetometer, while the field values may have
changed with T according to the phase diagrams in Fig.9.

After the scaling of the two data sets, and after additional
calibration to the measurements on the large mass 13.55 mg piece,
the locally saturated magnetization values were determined. The
criterium used for determining the magnetization for each state
($M_{j}$), was the onset $M(H)$ value (fig.12), \textit{i.e.} the
intersection of the linear fit of the $M_{j2}$ magnetization
plateau and the highest-slope linear fit of the $M(H)$ curve
during the $M_{j_1}$ to $M_{j_2}$ transition. More attention was
given to determining the magnetization for the first state
($M_{1}$), due to the limited field range over which this state
exists. Several criteria tried in this case (onset value, midpoint
between transitions, minimum in $dM/dH$ or midpoint on the
appropriate linear region on the $M(H)$ curves) resulted in almost
identical angular dependencies of $M_{1}$; moreover, using any of
the aforementioned criteria, we were still unable to follow this
state in the $M(H)$ curves for angles beyond $25^{0}$.

For $\theta~>~12^{0}$, the similar two sets of measurements are
shown in Fig.11b. The same criteria were used for the
determination of $M_{j}(\theta)$ and $H_{ci,j}(\theta)$. The two
lower metamagnetic transitions can also be seen in this region,
while of the higher two, only $H_{c2,3}$ is within our field
range; as the angle increases, the first two transitions move
closer in field ($H_{cAF,1}$ increases, while $H_{c1,2}$ doesn't
vary significantly with $\theta$), such that the $M_{1}$ state
becomes very narrow, making its determination very difficult.
$M_{2}$ and $M_{3}$ however appear as well defined plateaus,
continuously decreasing, and increasing respectively, from the
local extremum values seen at $\theta~=~0^{0}$. $H_{c2,3}$ has a
minimum of 38.9 kG around $\theta~=~30^{0}$.

The magnetization curves revealed four metamagnetic states, and
their angular dependence is presented in Fig.13b: $M_{1}$, $M_{2}$
and $M_{CL-SPM}$, which have local maxima at $\theta~=~0^{0}$
around $1.08~\mu_{B}/$Tb, $3.00~\mu_{B}/$Tb, and $6.25~\mu_{B}/$Tb
respectively, and $M_{3}$ which exists only beyond
$\theta~=~8^{0}$ and has a maximum of $5.06~\mu_{B}/$Tb at
$\theta~=~30^{0}$. Similar to the $H_{ci,j}$ in Fig.13a, the open
symbols in Fig.13b represent reflections of the measured data
across the $\theta~=~0^{0}$ (easy axis) direction.

The dotted lines in Fig.13a and b are fits to $H_{ci,j}(\theta)$
and $M_j(\theta)$ respectively, as calculated based on the model
that will be discussed below. Their angular dependencies are
described by $1~/~cos(\theta~\pm~\varphi_j)$, and
$cos(\theta\pm\varphi_j)$ respectively, with $\varphi_j~=~0^0$,
$30^0$ or $60^0$. These values are integer or half-integer
multiples of $360^0~/~n$, where $n~=~6$ in our hexagonal system.
Considering the six-fold symmetry of the this compound, these
simple geometrical relationships render TbPtIn as very similar to
RNi$_2$B$_2$C\cite{can02, can03} or DyAgSb$_2$\cite{mye03},
tetragonal compounds where the analogues $\varphi_j$ values were
$0^0$, $45^0$ or $90^0$ (integer or half-integer multiples of
$360^0~/~n$, where $n~=~4$).

As seen earlier in the cases of the tetragonal compounds
HoNi$_2$B$_2$C\cite{can02} or DyAgSb$_2$\cite{mye03}, simple
angular dependencies of the critical fields, as well as of the
locally saturated magnetizations exist in the hexagonal compound
TbPtIn; this will be further confirmed by similar geometrical
relationships that appear to exist in TmAgGe.

\section{data analysis}

The field and temperature dependent magnetization measurements on
TbPtIn (fig.2 and 5) have shown that this compound is extremely
anisotropic, with the magnetic moments confined to the hexagonal
basal plane. Moreover, when the direction of the applied field is
varied within the basal plane, six fold anisotropy of the
saturated magnetization is revealed in both TbPtIn and its
dilution (Tb$_x$Y$_{1-x}$)PtIn (Fig.6). Consequently, detailed
magnetization measurements with $H \bot c$ were performed,
allowing us to quantitatively describe the angular dependencies of
the critical fields $H_{ci,j}$ and locally saturated
magnetizations $M_j$ (Fig.13). By analogy to the four-position
clock model \cite{can02,mye03,kal04} for tetragonal systems, we
are now proposing a simple model for the net distribution of
moments in the hexagonal compound TbPtIn: three co-planar
Ising-like systems, $60^0$ apart in the basal plane. Such a
hypothesis was first suggested by the high field magnetization
values observed in the pure compound TbPtIn, as well as in the
highly diluted (Tb$_x$Y$_{1-x}$)PtIn. As the maximum measured
magnetization for TbPtIn was around $6~\mu_B/$Tb$^{3+}$ (far
smaller than the calculated $9~\mu_B$ value), it is reasonable to
assume the existence of more metamagnetic transitions beyond our
maximum applied field $H~=~140$ kG. However, in the highly diluted
compound, where, within our field and temperature ranges, we are
only probing the paramagnetic state, the magnetization also
reaches only $\sim 6~\mu_B/$Tb$^{3+}$ at the highest H. This is
consistent with the $M~=~6~\mu_B/$Tb$^{3+}$ corresponding to a
crystal-field limited saturated paramagnetic (CL-SPM) state.
Consequently we chose our model based on three Ising-systems such
that it described the hexagonal symmetry of the compound having
three magnetic ions in orthorhombic point symmetry, with the above
value corresponding to saturation (in the limit of high CEF
energy). In order to verify this hypothesis, the expected angular
dependencies of possible moment configurations resulting from such
a model will be compared with our measurements. Furthermore, we
will use our experimental results to refine the model, by
considering multiples of the three Ising-like systems, resulting
in more complex angular dependencies of the calculated
magnetization and critical field values.

In the $P\overline{6}2m$ space group, TbPtIn assumes a hexagonal
crystal structure, with 3 Tb$^{3+}$ ions at equivalent 3g
(orthorhombic) sites. The fact that a strong CEF anisotropy
confines the local moments to the basal plane calls for a two
dimensional model, greatly simplifying the analysis. (A schematic
description of an equivalent three-dimensional model has been
introduced for DyAgGe\cite{mor05}). Having three equivalent
magnetic moments in orthorhombic point symmetry, one possible way
to achieve the overall hexagonal symmetry is by restricting the
moments to three of the six-fold symmetry axes, $60^{0}$ apart,
while allowing for both the 'up' (solid arrows) and 'down' (dotted
arrows) positions for a given direction (Fig.14). Any specific
Tb-site would, at low temperatures, behave like an Ising system,
with each third of the sites having parallel Ising directions.
Thus, each metamagnetic state of TbPtIn can be described by a
multiple S of three Ising-like systems along three [120]
equivalent directions (the easy axes for this system). We will use
$\nwarrow$ , $\uparrow$ and $\nearrow$ symbols, to denote the
orientation of the three moments in their 'up' positions, and
$\searrow$ , $\downarrow$ and $\swarrow$ symbols respectively, for
the corresponding 'down' positions. The order of the arrows is not
meaningful for our model; only the number of arrows for each
orientation is significant for the net distribution of moments.
Moreover, we describe each metamagnetic state with the minimum-S
value moment configuration consistent with the experimental data.
However, higher S values are possible for most of the states, and
information about the wave vectors (\textit{e.g.} from scattering
experiments) would be required to determine unique S values.

Since in our experiments we only measure the projection of the
magnetic moment along the field direction, the angular dependence
of the magnetization $M_{j}$ per moment of an arbitrary 3
S-moments configuration is

$M_{j}(\theta)/\mu_{sat}($Tb$^{3+})~=~
\frac{1}{3*S}~[~\sum_{i=1}^Sm_{i}*\cos(\theta-60^0)+\sum_{i=1}^{S}m_{i}*\cos\theta~+~\sum_{i=1}^{S}m_{i}*\cos(\theta+60^0)~]$

where $\theta$ is a continuous variable representing the angle
between the applied field and the closest easy axis
($-30^0~\leq~\theta~\leq~30^0$), and the three sums give the
magnetization value due to each of the three directions of the
Ising-like systems; the $m_{i}$ parameters equal $\pm~1$,
depending on whether a certain moment is in the 'up' ($+~1$) or
'down' ($-~1$) position for the respective direction. We restrict
our model description to the $0^0~\leq~\theta~\leq~30^0$ angular
region, which, by symmetry across the $\theta~=~0^0$ direction,
also describes the $-30^0~\leq~\theta~\leq~0^0$ region.

For $S~=~1$, our model corresponds to one set of three such
Ising-like systems.  We assume that in high applied fields, the
three magnetic moments occupy the three allowed easy axes closest
to the direction of the field; as the field is lowered, the
metamagnetic transitions occur such that the measured
magnetization is being decreased with H. In this hypothesis, there
are three distinct moment configurations for the system:
($\nwarrow\uparrow\nearrow$) for the CL-SPM state,
($\searrow\uparrow\nearrow$) for intermediate field values, and
($\searrow\uparrow\swarrow$) for the AF ground state. The above
formula yields the following angular dependencies of the resulting
longitudinal (measured) magnetizations: $2/3*\cos\theta$,
$2/3*\cos(\theta-60^0)$ and $0$ respectively, represented by open
circles in Fig.15.

Fig.13b shows that such a model only describes the CL-SPM state
$M_{CL-SPM}(\theta)/\mu_{sat}($Tb$^{3+})~=~2/3*\cos\theta$ and the
AF ground state $M_{AF}/\mu_{sat}($Tb$^{3+})~=~0$ of TbPtIn;
according to the proposed model, the local moment configurations
from which the above angular dependencies follow are
($\nwarrow\uparrow\nearrow$) and ($\searrow\uparrow\swarrow$)
respectively. It is worth noting that the CL-SPM magnetization
value, calculated based on the above moment configuration, is
$6~\mu_B$, smaller than the measured $6.25~\mu_B$. One possible
explanation is that with increasing field, the system is slowly
approaching the CEF splitting energy. This is also consistent with
the increasing plateaus in the high field magnetization data in
Fig.5a; however, the extrapolation of these plateaus down to
$H~=~0$ results in smaller values for the [120] direction
($6.13~\mu_B$) and the [110] direction ($5.35~\mu_B$), closer to
the calculated values ($6.0~\mu_B$ and
$\sqrt{3}/2*6~\mu_B~=~5.2~\mu_B$ respectively).

To characterize all the other observed metamagnetic states, larger
S-values are needed, \textit{i.e.} the local moment configurations
are described by an integer multiple $S~>~1$ of sets of three
Ising-like systems. Fig.15 (crosses) also shows all possible
angular dependencies of the magnetizations resulting from such a
generalized model when $S~=~2$. By comparison with the
experimental data, it appears from Fig.13b that two more
metamagnetic states can now be described with $S~=~2$:

$M_3(\theta)/\mu_{sat}($Tb$^{3+})~=~2\sqrt{3}/6*\cos(\theta-30^0)$
(moment configuration
($\nwarrow\nwarrow\uparrow\uparrow\swarrow\nearrow$)~),

and

$M_2(\theta)/\mu_{sat}($Tb$^{3+})~=~2/6*\cos\theta$ (moment
configuration
($\searrow\nwarrow\uparrow\uparrow\swarrow\nearrow$)~) .

There is still one more metamagnetic state, $M_1$, which cannot be
described within the $S~=~2$ model; however, for most of the
angular range, its magnetization has an angular dependence
consistent with:
$M_1(\theta)/\mu_{sat}($Tb$^{3+})~=~2/18*\cos\theta$. The
$\cos\theta$ dependence (\textit{i.e.} $\cos$ is an even function)
requires that the moment configuration be symmetric with respect
to the $\theta~=~0^0$ direction. The simplest possibility is
($\nwarrow\uparrow\nearrow$), for which the magnetization varies
as $2/3*\cos\theta$; to this, a number of sets of three moments
needs to be added, with zero net magnetization (\textit{e.g.}
multiples of ($\searrow\uparrow\swarrow$) or
($\searrow\nwarrow\uparrow\downarrow\swarrow\nearrow$)), to get a
resulting magnetization amplitude of $2/18$. Consequently a
minimum $S~=~6$ configuration
($\searrow\searrow\searrow\searrow\searrow\nwarrow\uparrow\uparrow\uparrow\uparrow\uparrow\uparrow\swarrow\swarrow\swarrow\swarrow\swarrow\nearrow$)
or
($\searrow\searrow\searrow\nwarrow\nwarrow\nwarrow\downarrow\downarrow\uparrow\uparrow\uparrow\uparrow\swarrow\swarrow\swarrow\nearrow\nearrow\nearrow$)
yields the desired calculated magnetization $2/18*\cos\theta$.

Assuming the above net distributions of moments for the observed
metamagnetic states, one can derive the expected angular
dependencies of the critical fields. Comparison between the data
in Fig.13a and these calculated $H_{ci,j}(\theta)$ values will
further confirm or refute the net distributions of moments
proposed above.

Since the energy associated with a magnetic moment
$\overrightarrow{M}$ in an applied field $\overrightarrow{H}$ is
$\overrightarrow{M}\cdot\overrightarrow{H}$\cite{mye03}, the
corresponding energy difference $\Delta E_{ji}$ between
metamagnetic states $\overrightarrow{M_{i}}$ and
$\overrightarrow{M_{j}}$ is:

$\Delta
E_{ji}~=~\overrightarrow{M_{j}}\cdot\overrightarrow{H}-\overrightarrow{M_{i}}\cdot\overrightarrow{H}$

If there is a critical energy $E_{c}=\Delta E_{ji}$ to be exceeded
for a metamagnetic transition between states i and j to occur,
than the critical field value is given by:

$H_{ci,j}~=~\frac{E_{c}}{M_{j}-M_{i}}$,

where $M_{j}$ and $M_{i}$ are the measured (projections along the
field) respective magnetizations. The numerator in the above
expression is angle and field independent, and the angular
dependence of $H_{ci,j}$ follows only from the denominator. In
other words,

$H_{ci,j}~\sim~\frac{1}{M_{j}-M_{i}}$.

Consequently, the expected critical field values, shown as dotted
lines in Fig.13a, are:

$H_{cAF,1}(\theta)~\sim~1/\cos\theta$,

$H_{c1,2}(\theta)~\sim~1/\cos\theta$,

$H_{c2,3}(\theta)~\sim~1/\cos(\theta-60^{0})$,

$H_{c3,CL-SPM}(\theta)~\sim~1/\cos(\theta+60^{0})$

and $H_{c2,CL-SPM}(\theta)~\sim~1/\cos\theta$.

The reflections across the $\theta~=~0^0$ direction result from
the above formulas, when substituting $\theta$ with $-\theta$;
moreover, since $cos$ is an even function, this is equivalent to a
change in sign only for the $\varphi$ in the above expressions
written as $1/\cos(\theta-\varphi)$.

As described above and similar to the analogous study in the
tetragonal compound HoNi$_2$B$_2$C\cite{can02}, in most cases we
used maxima in $dM/dH$, and not the on-set criterion, to determine
the critical field values, because the magnetizations during the
transition were not always linear; however, comparison with the
calculated critical fields based on the above model is still
appropriate, given that only small departures from linearity were
encountered, mostly close to the bordering states ($M_i$ and
$M_j$). (The non-linear change of the magnetization with
increasing field indicates that other factors (\textit{i.e.} the
demagnetization factor of the sample\cite{suz06}, coexistence of
more than two phases, non-linear superposition of the various
states) may be responsible for the broadening of the transition).

Comparison of the measured critical fields and locally saturated
magnetizations (Fig.13a,b full and open symbols) with the
calculated values as described above (Fig.13a,b dotted lines)
confirms, in most cases, the assumed local distribution of
moments. However, the first metamagnetic state $M_1$ follows well
the calculated $2/18*\mu_{sat}($Tb$^{3+})*\cos(\theta)$ dependence
up to $\theta \approx 25^{0}$, after which it is difficult to
determine it with reasonable accuracy. Despite the fact that both
$H_{cAF,1}$ and $H_{c1,2}$ should depend on this magnetization
value, the former follows the expected angular dependence fairly
well, whereas the latter falls under the calculated $1/\cos\theta$
curve. Even though at this point we don't have a rigorous
calculation to support our assumption, we anticipate that the
proximity of the two lowest transitions requires a calculation
with more than two coexisting phases, which may render a better
fit of the observed experimental data in this region. On the other
hand, the $H_{c2,3}$ and $H_{c3,CL-SPM}$ critical fields, which
have the most evident angular dependence, are well fitted by the
calculated functions based on the present model. It is thus
reasonable to assume that, whereas possibilities for refining the
model exist, in the simple form that we present here it describes
our system fairly well.

The polar plot in Fig.16 helps in understanding how the hexagonal
crystal structure of this compound is reflected in the angular
dependence of the metamagnetic phase transitions: similar to the
polar phase diagrams for tetragonal compounds
HoNi$_{2}$B$_{2}$C\cite{kal04} or DyAgSb$_{2}$\cite{mye03}, when
we plot $H_{c}*\sin\theta$ vs. $H_{c}*\cos\theta$, the phase
boundaries become straight lines, with slopes equal to either
$\pm~1/\sqrt{3}$ or $\infty$. These slopes correspond to
directions either parallel or perpendicular to the high symmetry
axes (\textit{i.e.} $[110]$ or $[120]$) in the hexagonal
structure, just as within the four position clock model the
corresponding phase lines were either parallel or perpendicular to
the tetragonal high symmetry axes ($[110]$ or $[010]$). As
described in Myers \textit{et al.}\cite{mye03}, the equations of
these straight lines in polar coordinates can be used to verify
the transitions already discussed: if we substitute the above
slope values in the general formula

$R(\theta)~=~a/(\sin\theta-b*\cos\theta)$,

 for a line with slope b, we get:

$H_{c}(\theta)~\sim~1/\cos\theta$ for $b~=~\infty$

$H_{c}(\theta)~\sim~1/\cos(\theta-60^{0})$ for $b~=~-1/\sqrt{3}$

                          or

$H_{c}(\theta)~\sim~1/\cos(\theta+60^{0})$ for $b~=~1/\sqrt{3}$,

which are consistent with the angular dependencies of the
transitions determined above.

For the most part, the experimental points fall onto the
calculated straight lines, as expected. Some deviations from the
straight lines can be noticed, with the most evident one for
$H_{c1,2}$, for which we already emphasized the necessity of a
more complex model. In a similar manner (even though only for
fewer angles), $H_{c2,3}$ curves under the calculated straight
line as we move away from the [120] easy axis, while some even
smaller deviations from linearity are apparent in $H_{cAF,1}$;
this may indicate that special attention needs to be paid in
determining the angular dependence of the critical fields, when
rotating from the proximity of one easy axis to another.

\section{T\lowercase{m}A\lowercase{g}G\lowercase{e}}

We already reported the basic magnetic properties of
TmAgGe\cite{mor05}, which strongly resemble those of TbPtIn: the
magnetic susceptibility is extremely anisotropic (Fig.17),
indicating antiferromagnetic order below $T_{N} = 4.2$ K. The
local magnetic moments are confined by the strong CEF anisotropy
to the basal plane, both below and above $T_{N}$. This can also be
seen in the field dependent magnetization measurements, shown in
Fig.18, where in-plane anisotropy of the ordered state is also
apparent. Similar to the case of TbPtIn, several metamagnetic
transitions exist for both $H||[110]$ and $H||[120]$; these result
in magnetization values of $4.92~\mu_{B}$ and $4.30~\mu_{B}$
respectively, at $H~=~70$~kG, far below
$\mu_{sat}($Tm$^{3+})~=~7.0~\mu_{B}$, whereas for the c direction,
the magnetization is linear and much smaller up to the maximum
applied field. The ratio of the two in-plane magnetizations is
$\frac{M([120])}{M([110])}~=~0.87$, close to the $\cos30^{0}$
value expected within the model described before for TbPtIn for
the CL-SPM state. However, the two absolute values are larger than
the corresponding ones, calculated from the above model:
$M([110])~=~4.67\mu_{B}$ and $M([120])~=~4.00~\mu_{B}$, but the
extrapolation of the high-field plateaus down to $H~=~0$ (solid
lines in Fig.18) yields magnetizations very close to these
calculated values. As in the case of TbPtIn, the slight increase
of the magnetization plateaus after the supposed saturation may be
caused by the slow approach of the CEF splitting energy.

In calculating the above expected magnetization values, we assumed
the easy axes to be along the [110]-equivalent directions, based
on the directions where maximum magnetization values at
$H~=~70$~kG were achieved (fig.18). This is consistent with the
angular dependent magnetization measurement for $H~=~70$~kG shown
in Fig.19, where angle $\theta$ was measured from the [110]
direction; thus the six-fold symmetric magnetization has maxima
occurring for the $[110]$-equivalent directions (\textit{i.e.} for
$\theta~=~n*60^{0}$, where \textit{n} is an integer).
Consequently, for TmAgGe the angle $\theta$ will be measured from
the closest [110] easy axis. The comparison of TmAgGe and TbPtIn
indicates that, even though the easy axes in the two compounds
correspond to the two different sets of six-fold symmetry
directions, as we shall see, their physical properties are very
similar.

Around each easy axis, these magnetization measurements follow the
$\cos\theta$ angular dependence (the solid line in Fig.19), as
expected within our proposed model. Some differences between the
experimental data (filled circles in fig.19) and the calculated
magnetization could be caused by small misalignment of the sample
(rendering slightly asymmetric measured peaks), or by the strong
interactions between the local moments. Similar to TbPtIn, this
also indicates that different metamagnetic states cross $H~=~70$
kG at different angles.

From the $~M(T)\mid_{H,\theta}$ (fig.20) and
$~M(H)\mid_{T,\theta}$ (fig.21) measurements, detailed $H-T$ phase
diagrams for this compound can be determined. They are shown in
Fig.22a,c, for field along the [110] or $\theta~=~0^0$, and [120]
or $\theta~=~30^0$ directions respectively, with an
intermediate-position $\theta~\approx~24^0$ phase diagram in
Fig.22b. In the same manner used for TbPtIn, the points in these
phase diagrams have been obtained from maxima in either
$d(M*T/H)/dT$ for constant field (full symbols) or in $dM/dH$ for
fixed temperatures (open symbols).

For $H||[110]$ (fig.22a), at low temperatures the
antiferromagnetic, AF, ground state exists for $H~\leq~3.1$ kG,
followed by a small intermediate phase $M_1$ (up to $\sim~4.4$ kG)
and a larger state $M_2$ above. This latest phase extends up to
$8.9$ kG, after which, at low temperatures, the system reaches the
crystal field-limited saturated paramagnetic, CL-SPM, state. As
temperature is increased, the $M_1$ phase disappears around $2.5$
K , and a direct transition from the $AF$ to the $M_2$ state
occurs at a decreasing critical field value. The upper phase
boundary (for the $M_2$ to the PM state transition) also falls
down in field as T increases, such that at very low fields only
one transition is observed close to $T_N~=~4.2$ K.

As we rotate away from the easy axis, the low-field phase diagram
changes very little, with a small enhancement of the critical
field values towards low temperatures. As field is being
increased, $M_2$ is getting smaller as a new distinct phase $M_3$
forms. Its upper bordering line appears to have a strong angular
dependence, as can be seen in Fig.22b and c, similar to the
upper-most phase boundary seen in TbPtIn : for $H||[120]$
(Fig.22c), this phase boundary is an almost vertical line at
$T~\approx~4.5$K, up to our maximum applied field $H~=~70$ kG. As
a consequence, the corresponding points on this line have been
determined from $d(M*T/H)/dT$ data, as shown in Fig.20c for high
fields, and could not be identified in the field-dependent
derivatives. At low temperatures, the $H_{cAF,1}$ and $H_{c1,2}$
values (3.62 kG and 4.86 kG respectively) are very close to the
corresponding ones in the $[110]$ direction, whereas the $M_2$ to
$M_3$ transition occurs around 7.0 kG. These three phase lines
merge around $T~=~3.0$ K, such that for higher temperatures a
single transition occurs at decreasing fields. This line appears
to intersect the $H = 0$ axis around $T_N~=~4.2$ K.

The intermediate-orientation phase diagram presented in Fig.22b
allowed us to observe the upper-most phase line moving down in
field at low temperatures, such that for $\theta~\approx~24^0$, it
intersects the $T~=~0$ axis close to 40.1 kG. In this orientation,
this phase boundary can be identified in the M(H) derivative, as
shown in Fig.23 for $T~=~1.85$ K. However, the high field peak in
$dM/dH$ is poorly defined, making the determination of the
corresponding critical field value more difficult. For a more
precise estimate, another criterium was used together with the
derivative maxima, as illustrated in the inset in Fig.23: the
mid-point (large dot) on the highest-slope linear fit (solid line)
of the magnetization data around the transition. Also shown is the
error bar for this critical field value, as determined from the
two criteria used here.

The fact that this line is now apparent in both $M(T)$ and $M(H)$
data is further confirmation that this phase boundary exists,
whereas at lower fields, the only noticeable difference from the
$H||[120]$ direction is the persistence of the $M_2$ state up to
higher (\textit{i.e.} $\sim~3.5$ K) temperatures.

A number of similarities between TbPtIn and TmAgGe have already
been established: same crystal structure, antiferromagnetic ground
state, extremely anisotropic magnetization, in-plane anisotropy
and metamagnetism leading to crystal field-limited saturated
magnetizations smaller than the calculated single ion $\mu_{sat}$
values. As a consequence, we proceed to study the angular
dependence of the planar metamagnetism in TmAgGe, and subsequently
apply the model developed for TbPtIn to the case of this compound.

When we fix the temperature at $T~=~2$ K, the angular dependence
of the metamagnetic transitions can be studied based on the $M(H)$
isotherms shown in Fig.24. The critical fields and the locally
saturated magnetization values (full symbols in fig.25) have been
determined as maxima in $dM/dH$, and from on-set values
respectively (see the TbPtIn section). An exception was made for
$M_{CL-SPM}$ above $10^0$, and the criterion used for determining
this state is described below. Moreover, because of the proximity
of the first two transitions, the $M_1$ state is poorly defined;
no precise saturated magnetization data could be extracted for
this state, but the phase diagrams in Fig.22, as well as the
angular dependent critical fields in Fig.25a, are consistent with
the existence of this phase. As before, the open symbols in Fig.25
represent reflections of the measured data across the
$\theta~=~0^0$ direction. The TmAgGe measurements allowed us to
determine the critical field and locally saturated magnetization
values for the full angular range ($-30^0~\leq~\theta~\leq~30^0$);
the resulting somewhat asymmetric data (most obvious in the case
of the $M_2$ data) maybe due to a small sample misalignment. The
experimental data (full symbols) together with the reflections
(open symbols) in Fig.25 give the caliper of the error bars for
these measurements.

For $\theta~\leq~10^{0}$, two closely spaced metamagnetic
transitions can be seen in Fig.25a, with critical fields
$H_{cAF,1}~=~3.0$ kG and $H_{c1,2}~=~4.37$ kG respectively at
$\theta~=~0^{0}$, followed by a third transition $H_{c2,CL-SPM}$
at $\sim~9.37$ kG. It should be noted that these $H_{ci,j}$ values
are slightly different from the corresponding ones (3.10 kG, 4.36
kG and 8.92 kG respectively at $T~=~2.0$ K, $\theta~=~0^0$) in the
$H-T$ phase diagrams (Fig.22a), as they have been determined from
two distinct measurements. Thus small errors in the angular
position ($\pm~1^0$) may convert into small errors in the critical
field values ($\leq~3\%$).

Somewhat larger differences between the two data sets are observed
for $\theta~\geq~10^0$, specifically for $H_{c3,CL-SPM}$, which
varies more rapidly with the angle than any other critical field.
In this angular region, the two lower transitions occur at almost
the same critical fields as below $10^0$, whereas the critical
field for the third one slowly decreases with angle, as a fourth
transition appears and rapidly moves up in field. Consequently,
the $M_1$ metamagnetic state changes little with the angle,
whereas the $M_2$ state narrows down as the bordering critical
fields move closer to each other. The fourth transition being very
broad makes the determination of the $M_3$ state fairly difficult.
Also, with $H_{c3,CL-SPM}$ broadening out and rapidly moving
towards our field limit (\textit{i.e.} 70 kG), it was difficult to
get a meaningful linear fit of the $M(H)$ curves during the $M_3$
to $CL-SPM$ transition; instead we used the intersection of the
maximum slope line corresponding to the $M_2$ to $M_3$ transition,
and the best linear fit of the highest magnetization state, to
determine $M_{CL-SPM}$ for $\theta~\geq~10^{0}$ (Fig.24b, inset).

The best fits to the experimentally measured angular dependent
data are shown in Fig.25 as dotted lines. We will use these fits
to infer the net distribution of moments as multiples S of three
Ising-like systems, similar to the case of TbPtIn. As mentioned
before, we infer that the $M_1$ state should exist based on the
angular dependent critical fields in Fig.25a, and the $T~=~2.0$ K
metamagnetic phases revealed by the phase diagrams in Fig.22.
Consequently, in Fig.25b we are only showing the expected angular
dependence of such a phase, by analogy with the TbPtIn case:
$M_1^{calc}/\mu_{sat}($Tm$^{3+})~=~2/18*cos\theta$, which appears
to be the upper limit of these magnetization values, as indicated
by the error bars shown in Fig.25b. As already seen for TbPtIn,
the moment configuration that would result in such a $M_1$
magnetization is a $S~=~6$ state:
($\searrow\searrow\searrow\searrow\searrow\nwarrow\uparrow\uparrow\uparrow\uparrow\uparrow\uparrow\swarrow\swarrow\swarrow\swarrow\swarrow\nearrow$)
or
($\searrow\searrow\searrow\nwarrow\nwarrow\nwarrow\downarrow\downarrow\uparrow\uparrow\uparrow\uparrow\swarrow\swarrow\swarrow\nearrow\nearrow\nearrow$).
$M_2(\theta)$ has a maximum value at $\theta~=~0^{0}$ equal to
$2.37~\mu_{B}/$Tm$^{3+}$, close to $2/6*\mu_{sat}(Tm^{3+})~=~
2/6*7\mu_{B}/$Tm$^{3+}$, and a $\cos\theta$ angular dependence.
This suggests that a possible net distribution of moments for this
state, realized with a minimum $S~=~2$, could be
($\searrow\nwarrow\uparrow\uparrow\swarrow\nearrow$). If one local
moment is flipped from its $\swarrow$ position to $\nearrow$ in
the previous state, the resulting state could be described by the
($\searrow\nwarrow\uparrow\uparrow\nearrow\nearrow$)
configuration, whose magnetization varies as
$2\sqrt{3}/6*\mu_{sat}($Tm$^{3+})*\cos(\theta-30^{0})$; this fits
well the measured $M_3$ data, indicating that the previously
assumed local moment distribution may be appropriate for this
metamagnetic state. When all the magnetic moments are in their
'up' positions, the CL-SPM
($\nwarrow\nwarrow\uparrow\uparrow\nearrow\nearrow$) state is
achieved, and the corresponding angular dependence is
$4/6*\mu_{sat}($Tm$^{3+})*\cos\theta$. This best describes the
last observed metamagnetic state, which has a maximum of
$4.68\mu_{B}/$Tm$^{3+}$ around $\theta~=~0^{0}$, very close to
$4/6*\mu_{sat}($Tm$^{3+})~=~4/6*7\mu_{B}/$Tm$^{3+}$. (As before,
the order of the arrows used to describe the net distribution of
moments has no physical meaning).

According to the calculation given in the case of TbPtIn, for
TmAgGe one would also expect the critical fields to vary with the
angle $\theta$ as

$H_{ci,j}~\sim~\frac{1}{M_{j}-M_{i}}$.

Using the net distributions of moments assumed above to best
describe the locally saturated magnetization states, we expect the
following angular dependencies of the critical fields:

$H_{cAF,1}(\theta)~\sim~1/\cos\theta$,

$H_{c1,2}(\theta)~\sim~1/\cos\theta$

$H_{c2,3}(\theta)~\sim~1/\cos(\theta-60^{0})$,

$H_{c3,CL-SPM}~\sim~1/\cos(\theta+60^0)$

and $H_{c2,CL-SPM}(\theta)~\sim~1/\cos\theta$.

The experimental data (full symbols in Fig.25a) are in good
agreement with these calculated critical fields, with
$H_{c2,CL-SPM}(\theta)$ present only for $\theta~\leq~10^{0}$,
while the $H_{c2,3}$ and $H_{c3,CL-SPM}$ exist only for
$\theta~\geq~10^{0}$. This is consistent with the presence of the
$M_3$ state for angles larger than $10^{0}$, even though
experimentally we were only able to accurately determine it for
$\theta~\geq~16^0$. (The bordering transitions of this state are
very close in field when $\theta$ is close to $10^{0}$, which made
the determination of $M_3$ in this angular region difficult). It
is worth pointing out the excellent fit of the measured
$H_{c3,CL-SPM}$ data with the calculated angular dependency, this
transition showing the most dramatic change with the angle
$\theta$.

Apart from the absolute values of the critical fields and the
locally saturated magnetizations, the $H_c(\theta)$ and
$M_{sat}(\theta)$ phase diagrams for TmAgGe (Fig.25) are identical
to the TbPtIn analogues in Fig.13: to the same number of critical
fields with identical angular dependencies correspond identical
metamagnetic values (scaled to the saturated moment of the
respective R$^{3+}$ ion), which also vary similarly with the
angle. This is consistent with our model being indeed a general
description of the Fe$_2$P-type systems, or even more generally,
of hexagonal systems with the R in orthorhombic point symmetry.

When the $H_c(\theta)$ phase diagram for TmAgGe is converted into
a polar plot (Fig.26), similar to that for TbPtIn, we again notice
that the phase boundaries are straight lines, with
$\pm~1/\sqrt{3}$ or $\infty$ slopes. From the equations of these
straight lines, we can once more confirm the transitions
determined before:

$H_{c}(\theta)~\sim~1/\cos\theta$ for $b~=~\infty$

$H_{c}(\theta)~\sim~1/\cos(\theta-60^{0})$ for $b~=~-1/\sqrt{3}$

                          or

$H_{c}(\theta)~\sim~1/\cos(\theta+60^{0})$ for $b~=~1/\sqrt{3}$.

As already noted, in the case of TmAgGe, we were able to determine
the critical fields from experimental data for
$\theta~=~-30^{0}~..~30^{0}$, as seen in the phase diagrams in
Fig.25a, as well as in Fig.26. Slight differences between the
expected straight lines in Fig.26 and measured critical fields can
be noticed for $H_{c2,3}$ for angles close to $\pm~30^{0}$, or for
$H_{c3,CL-SPM}$ also for large angles; besides being a consequence
of small misorientation of the sample, this may indicate, similar
to the TbPtIn case, that a more complex model needs to be used to
describe the regions around the 'hard' in-plane direction.

\section{Angular dependent metamagnetism at $T~=~20$ K in T\lowercase{b}P\lowercase{t}I\lowercase{n}}

Whereas TmAgGe has fairly simple $H-T$ phase diagrams, with all
metamagnetic phases present at low temperatures, the TbPtIn phase
diagrams are somewhat more complex, manifesting an additional
intermediate-temperature phase, $M_4$ (Fig.9). In order to perform
a similar angular dependent study of this metamagnetic state, a
set of $M(H)\mid_{\theta}$ data was taken at $T~=~20$ K. Assuming
that all existing transitions have been identified and are shown
in fig.9 (at least below 20 K), at this temperature the $M(H)$
curves should intersect the same magnetic phases as in the low
temperature case, with the exception of $M_1$; instead, the
measurements at $T~=~20$ K intersect the bubble-like phase $M_4$,
as seen in the three different orientations phase diagrams in
Fig.9.

Fig.27 shows the $M(H)$ isotherms at $T~=~20$ K for various angles
$\theta$. The $M_{j}(\theta)$ and $H_{ci,j}(\theta)$ phase
diagrams have been determined as described before for TbPtIn or
TmAgGe for $T~=~2$ K, and are shown in fig.28. It should be noted
that, due to the enhanced temperature, all transitions are
broadened, and the locally saturated magnetization plateaus are no
longer horizontal. Both of these facts make the analysis of these
data somewhat harder and more ambiguous.

For $\theta~\leq~12^0$ (Fig.28a), the lowest transition changes
very little with angle, having a critical field value $H_{c
AF,4}~\approx~8$ kG. As field is being increased, two more
transitions can be observed for angles lower than $8^0$, with
local minima of the critical fields, at $\theta~=~0^0$, of $H_{c
4,2}~=~16.2$ kG and $H_{c 2,CL-SPM}~=~48.7$ kG respectively. For
larger angles, the highest transition splits into two different
ones, $H_{c 2,3}$, with decreasing values as we rotate away from
the easy axis, and $H_{c 3,CL-SPM}$, which rapidly increases above
our field limit (\textit{i.e.} 55 kG) around $\theta~=~12^0$. It
should be noted that we are still referring to the high-field
state at $T~=~20$ K as the crystal field-limited saturated
paramagnetic CL-SPM state, even though it is possible that
cross-over to the paramagnetic PM state has occurred between 2 K
and 20 K at high H. (This would be a plausible explanation for the
measured magnetization values for this high field state being, as
seen below, lower than the calculated values.)

The locally saturated magnetization of the $M_4$ state is equal to
$\sim~0.25~\mu_B/$Tb at $\theta~=~0^0$ and doesn't appear to
change much with the angle. According to the $H-T$ phase diagrams
in Fig.9, all higher metamagnetic states are identical at low
($T~=~2$ K) temperature and at $T~=~20$ K; consequently, they seem
to have similar angular dependencies (fig.28b): $M_2$ and
$M_{CL-SPM}$ have local maxima around $3.3\mu_B/$Tb and
$5.7\mu_B/$Tb respectively, at $\theta~=~0^0$, and slowly decrease
with increasing angle in this region. Beyond $\sim~8^0$, a third
metamagnetic state should exist, defined by the $H_{c 2,3}$ and
$H_{c 3,CL-SPM}$ critical fields; however, it is difficult to
identify it in this angular region, given the broadness of the
bordering transitions and the $H_{c 3,CL-SPM}$ proximity to our
field limit.

As we move further away from the easy axis (\textit{i.e.}
$\theta~>~12^0$, Fig.28b), the lower two metamagnetic states can
again be observed, whereas the CL-SPM state may still exist for
fields larger than 55 kG (also apparent from the $H-T$ phase
diagram in Fig.9b). Also, we can now see the third state $M_3$
slowly increasing with angle $\theta$, similar to the low
temperature case.

The $T~=~20$ K phase diagrams (fig.28) are very similar to their
low temperature analogues (Fig.13), except for the $M_4$ state,
and some evident differences between the experimental data and the
model calculations (dotted lines). It was rather difficult to
determine the angular dependencies of $M_4$ and $M_2$, therefore
we had to infer the possible theoretical fits in a more indirect
way: as we already mentioned, the state described by $M_2$ at
$T~=~20$ K should be the same as the corresponding one at low
temperature, since they characterize the same metamagnetic phase.
Therefore we expect it to vary with the angle like
$2/6*\mu_{sat}($Tb$^{3+})*\cos\theta$ (fig.13b). For large angles,
this is consistent with the measured data in Fig.28b, whereas
significant deviations can be noticed closer to $\theta~=~0^0$.
The lower magnetization state $M_4$ has much smaller values than
any of the states characterized at $T~=~2$ K, therefore we cannot
fit it accurately with a calculated angular dependence; however,
$H_{c4,2}$ can be fitted with 16.2 kG $*1/\cos\theta$ and should
relate to $M_4$ through

$H_{c4,2}(\theta)~\sim~1/[M_2(\theta)-M_4(\theta)]$

or

$1/\cos\theta~\sim~
1/[~2/6*\mu_{sat}($Tb$^{3+})*\cos\theta-M_4(\theta)]$.

Thus $M_4(\theta)$ should vary like $M_{sat,4}*\cos\theta$, with a
locally saturated magnetization $M_{sat,4} \approx
0.25*\mu_{sat}($Tb$^{3+})$, but, as already mentioned, it is
difficult to determine it with reasonable accuracy. However, the
corresponding local moment configuration should be similar to the
low temperature one, in order to get the $cos\theta$ dependence,
except that the number S of three Ising-like systems which would
yield the appropriate $M_{sat,4}$ value is uncertain.

All other locally saturated magnetizations and critical fields can
be best fitted with the same angular dependencies as for the low
temperature case:

$M_{3}(\theta)~=~
2\sqrt{3}/6*\mu_{sat}($Tb$^{3+})*\cos(\theta-30^0)$,

$M_{CL-SPM}(\theta)~=~4/6*\mu_{sat}($Tb$^{3+})*\cos\theta$,

and

$H_{cAF,4}~\sim~1/\cos\theta$,

$H_{c2,3}~\sim~1/\cos(\theta-60^0)$,

$H_{c2,CL-SPM}~\sim~1/\cos\theta$,

$H_{c3,CL-SPM}~\sim~1/\cos(\theta+60^0)$.

Apart from the already mentioned differences between the measured
data and the calculated curves, small departures from the
corresponding theoretical angular dependencies can be noticed for
$M_3(\theta)$ and $H_{c2,3}$; a more significant difference
appears for the saturated magnetization state
$M_{CL-SPM}(\theta)$, which seems to have the expected angular
dependence, but with smaller values than the calculated ones. This
may be a high-temperature effect (\textit{i.e.} cross-over from
low-T CL-SPM state to high-T paramagnetic PM state), or it may be
one more indication that a more refined model is needed.

A polar plot analogues to the low temperature case (Fig.29) shows
that at $T~=~20$ K, the critical fields are still well described
by straight lines, but with more pronounced differences between
experiment and the theoretical calculations. We attribute these to
the thermal broadening at this temperature, but, as already seen
in the low temperature case, a requirement for a more complex
model cannot be excluded.

\section{Summary}

Motivated by the extensive work done on highly anisotropic local
moment systems with tetragonal unit cells and unique rare earth
sites of tetragonal point symmetry, we have performed detailed
studies on two highly anisotropic, local moment, hexagonal
compounds: TbPtIn and TmAgGe.  Whereas both of these compounds are
ternary members of the Fe$_2$P class of materials, they have
different ligands.  In addition, whereas both of these compounds
manifest extreme planar anisotropy, they have different easy axes:
$[120]$ for TbPtIn and $[110]$ for TmAgGe.  Even with these
differences we have found that these two compounds have very
similar $H-T$ as well as $H-\theta$ phase diagrams.

TbPtIn and TmAgGe have a single rare earth site with orthorhombic
point symmetry (with three rare earth sites per unit cell) and
both compounds have high field saturated moments well below the
single ion values.  These two observations, combined with our
experience with the four position clock model that was developed
for the tetragonal compounds with rare earths in tetragonal
symmetry, lead us to propose a similar model for these
Fe$_2$P-type compounds:  a triple coplanar Ising model, which
consists of three Ising-like moments per unit cell, with their
Ising axes within the basal plane and rotated by $60^0$ with
respect to each other. This model preserves the six-fold symmetry
at high fields and also explains why the saturated moments are
significantly lower than the free ion values. By analyzing the
magnitudes and angular dependencies of the critical metamagnetic
fields, as well as the locally saturated magnetizations within the
framework of this model, we can infer the net distribution of
moments along the six possible moment orientations. However,
field-dependent neutron diffraction or magnetic x-ray measurements
are needed to test these hypothetical net distributions of moments
and to obtain the propagation vectors for each magnetically
ordered state. (Some preliminary experiments on TbPtIn are already
reported\cite{gar12}, confirming the two low-field ordering
temperatures that we observed in the magnetization measurements.
Moreover, the neutron data are consistent with our inferred
direction of the moments.)

The successful extension of the four state clock model to the
current triple coplanar Ising-like model implies that a wider set
of local moment compounds with planar anisotropy can be understood
in a similarly simple manner. Clearly tetragonal unit cell
compounds with the rare earth in orthorhombic point symmetry could
be expected to behave in a manner similar to TbPtIn and TmAgGe,
\textit{i.e.} to form a class of double coplanar Ising model
materials. In a similar manner, hexagonal unit cell compounds with
the rare earth in hexagonal point symmetry could be expected to
behave in a manner similar to HoNi$_2$B$_2$C\cite{can02} or
DyAgSb$_2$\cite{mye03}, \textit{i.e.} to form a class of six
position clock model materials. Hexagonal unit cell materials may
offer one other, potentially new class of materials: highly planar
anisotropic compounds with the rare earth in trigonal point
symmetry.  In this case we anticipate a double coplanar three
position clock model. Such compounds would have two magnetic
sites, each with three possible positions $120^0$ apart. Fig.30
gives schematic representations of all expected five models
described above.

\section{acknowledgments}

We thank Lizhi Tan for helping index the macroscopic facets of our
crystals. We are also grateful for fruitful discussions with
Carsten Detlefs and Andreas Kreyssig. Ames Laboratory is operated
for the U.S. Department of Energy by Iowa State University under
Contract No. W-7405-Eng.-82. This work was supported by the
Director for Energy Research, Office of Basic Energy Sciences.

\clearpage

\begin{figure}
\begin{center}
\includegraphics[angle=0,width=150mm]{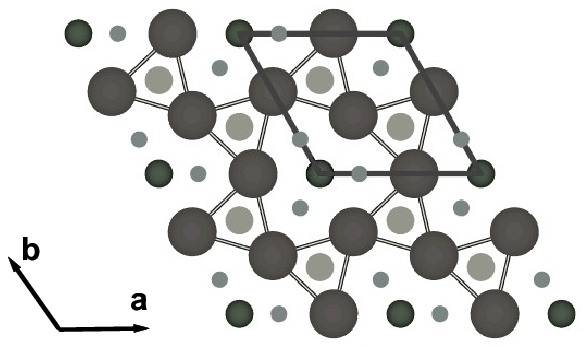}

\caption{ Projection of the \textit{ZrNiAl}-type crystal structure
along the hexagonal c-axis: R -large circles, M (Pt or Ag)-medium
circles, X (In or Ge)- small circles. Light circles: $z~=~0$
plane, dark circles: $z~=~1/2$ plane. (Note: the Pt or Ag atoms
(medium circles) appear in both planes.)}
\end{center}
\end{figure}

\clearpage

\begin{figure}
\begin{center}
\includegraphics[angle=0,width=150mm]{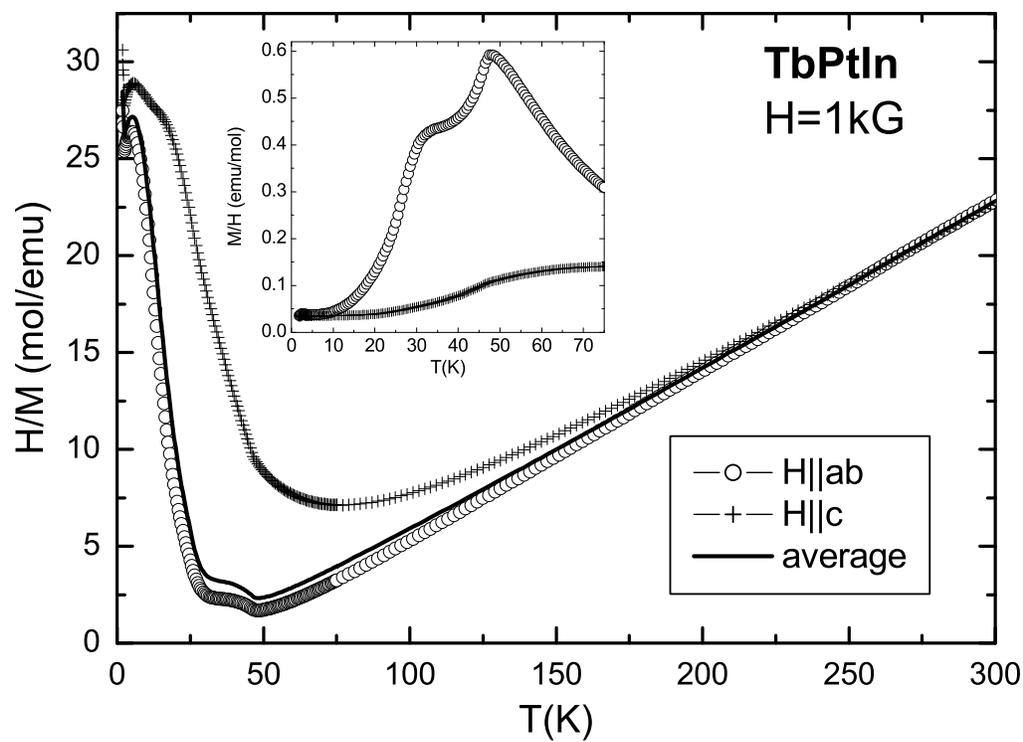}

\caption{Anisotropic inverse susceptibility of TbPtIn (symbols)
and the calculated average (line); inset: low-temperature
anisotropic susceptibilities.}
\end{center}
\end{figure}

\clearpage

\begin{figure}
\begin{center}
\includegraphics[angle=0,width=100mm]{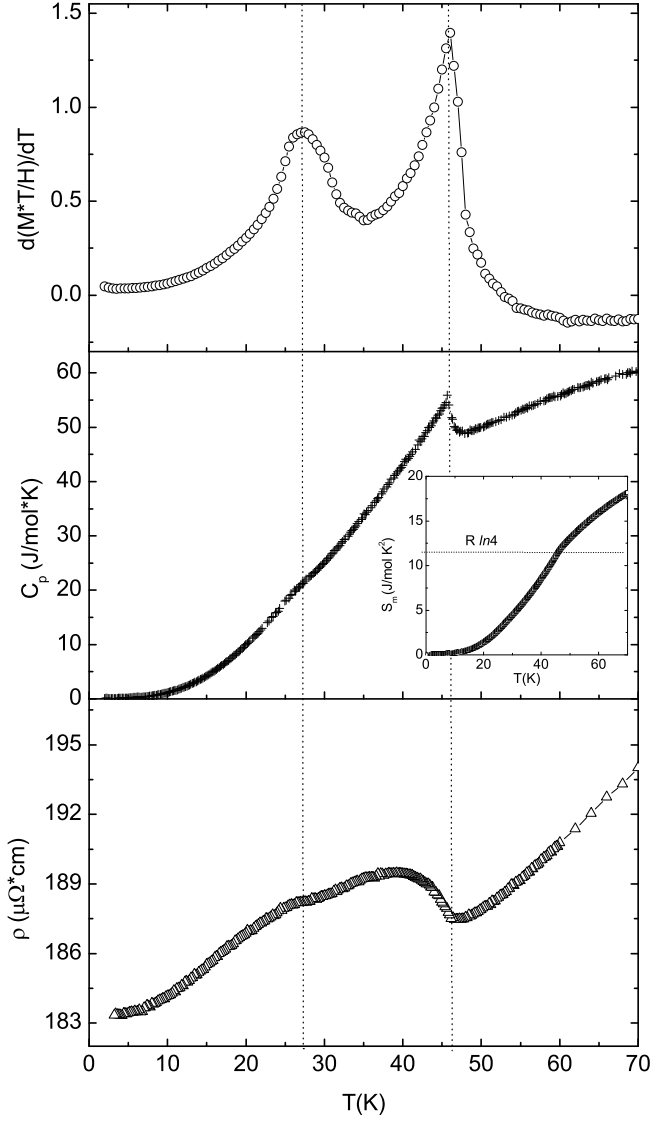}

\caption{(a) Low-temperature $d(\chi_{ave}*T)/dT$, with vertical
dotted lines marking the peaks positions; (b) specific heat
$C_p(T)$; inset: magnetic entropy $S_m(T)$; (c) low-temperature
resistivity $\rho(T)$, for current flowing in the basal plane ($i
\parallel ab$).}
\end{center}
\end{figure}

\clearpage

\begin{figure}
\begin{center}
\includegraphics[angle=0,width=150mm]{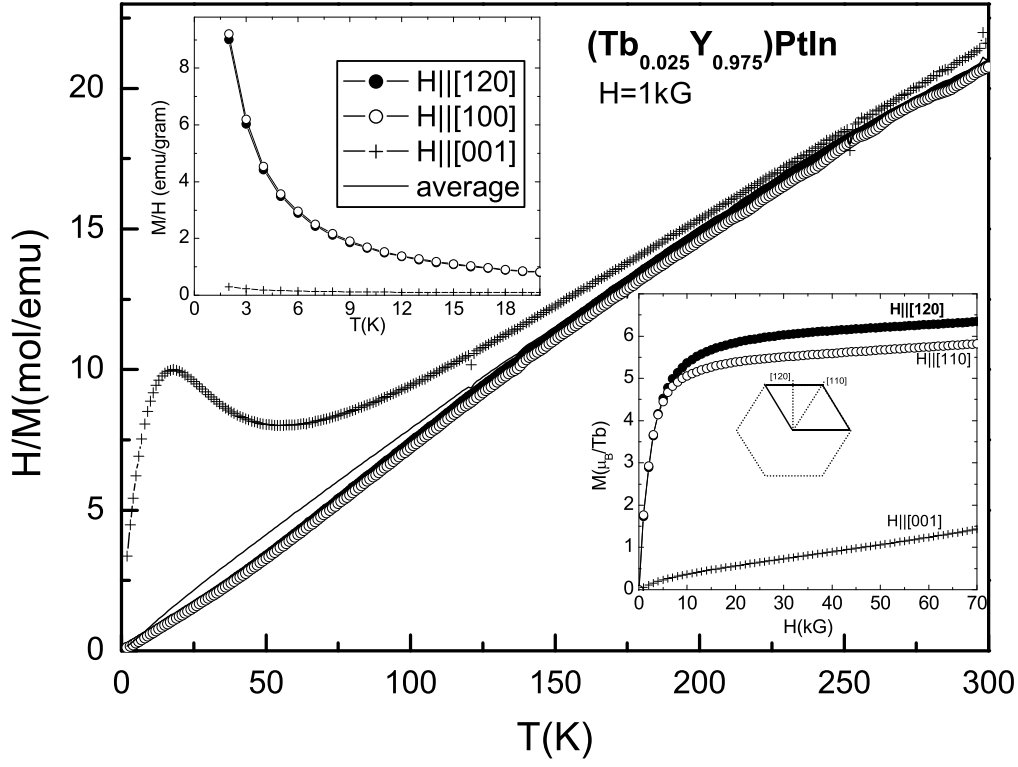}

\caption{Anisotropic inverse susceptibility (symbols) and
calculated average (line) for H = 1 kG; upper inset: low-
temperature anisotropic susceptibility; lower inset:
field-dependent anisotropic magnetization, for T = 2 K.}
\end{center}
\end{figure}

\clearpage

\begin{figure}
\begin{center}
\includegraphics[angle=0,width=130mm]{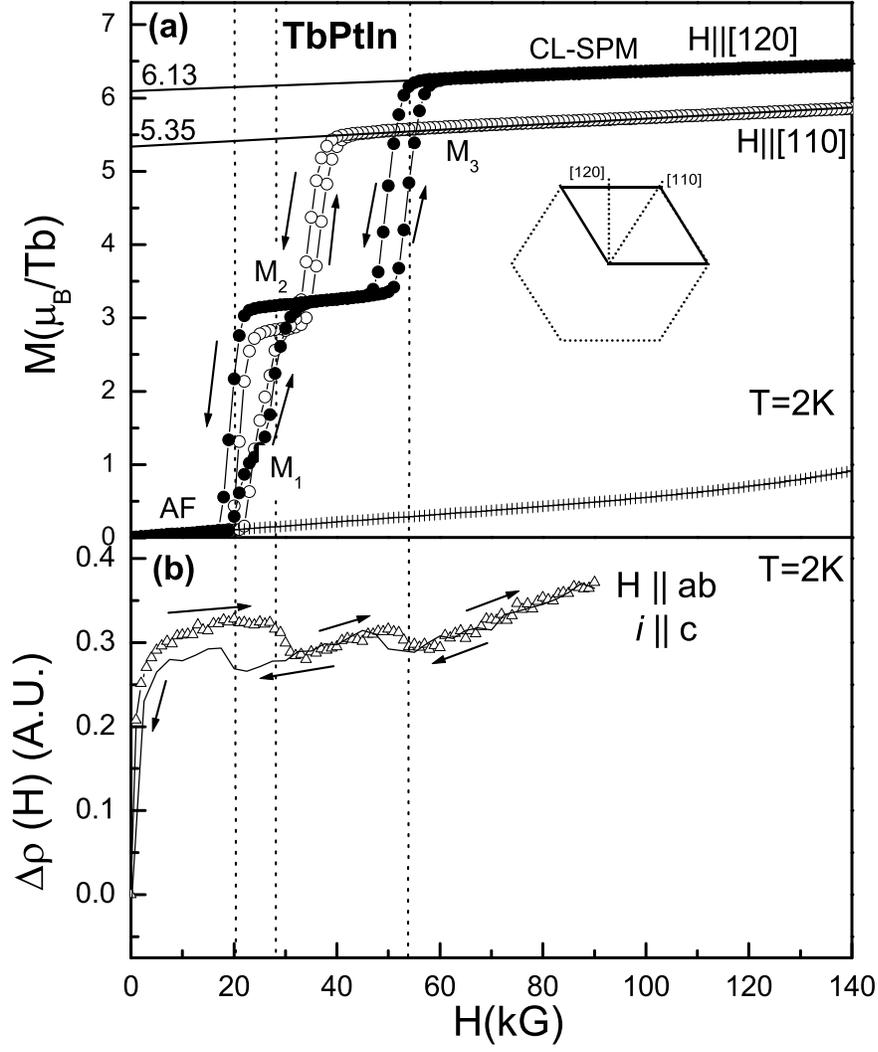}

\caption{(a) Anisotropic field-dependent magnetization (increasing
and decreasing field data indicated by arrows) and (b) transverse
magnetoresistance, for increasing (symbols) and decreasing (line)
field.}
\end{center}
\end{figure}

\clearpage

\begin{figure}
\begin{center}
\includegraphics[angle=0,width=150mm]{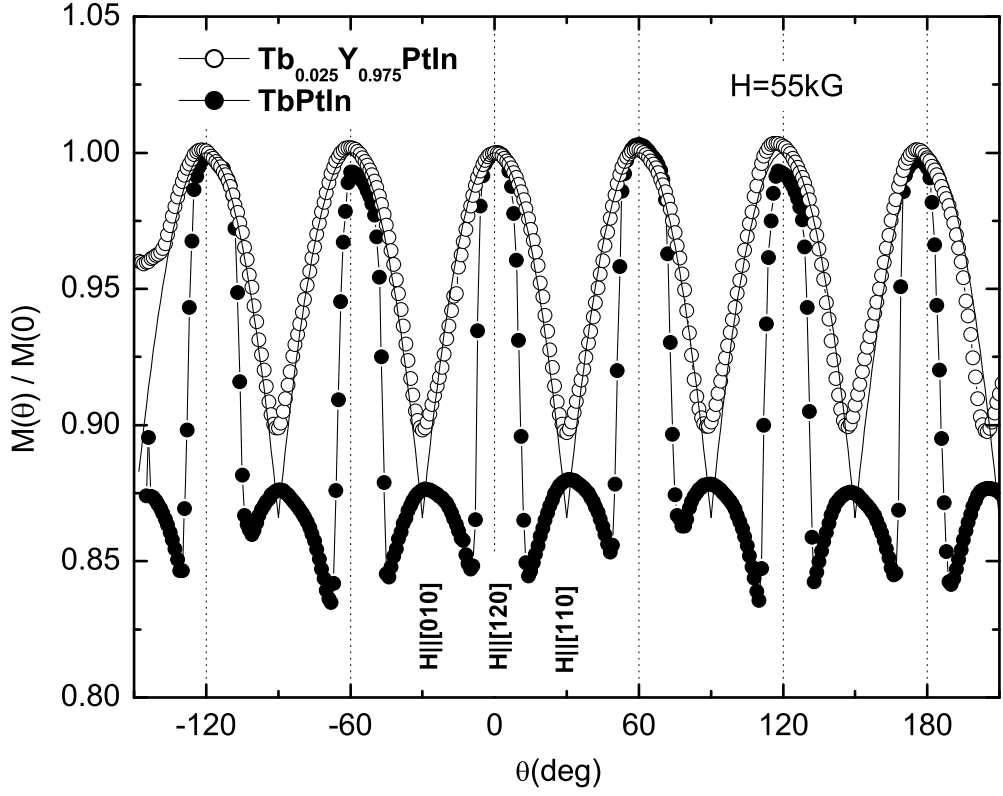}

\caption{ M($\theta$) of TbPtIn (full circles) and
Tb$_{0.025}$Y$_{0.975}$PtIn (open circles) for T = 2 K and H = 55
kG, H$\perp$c. Solid line represents the calculated
M$_{max}*\cos(\theta-n*60^0)$, \textit{n}-integer. Note:
$\theta~=~0^0 $ is defined at the [120] direction.}
\end{center}
\end{figure}

\clearpage

\begin{figure}
\begin{center}
\includegraphics[angle=0,width=130mm]{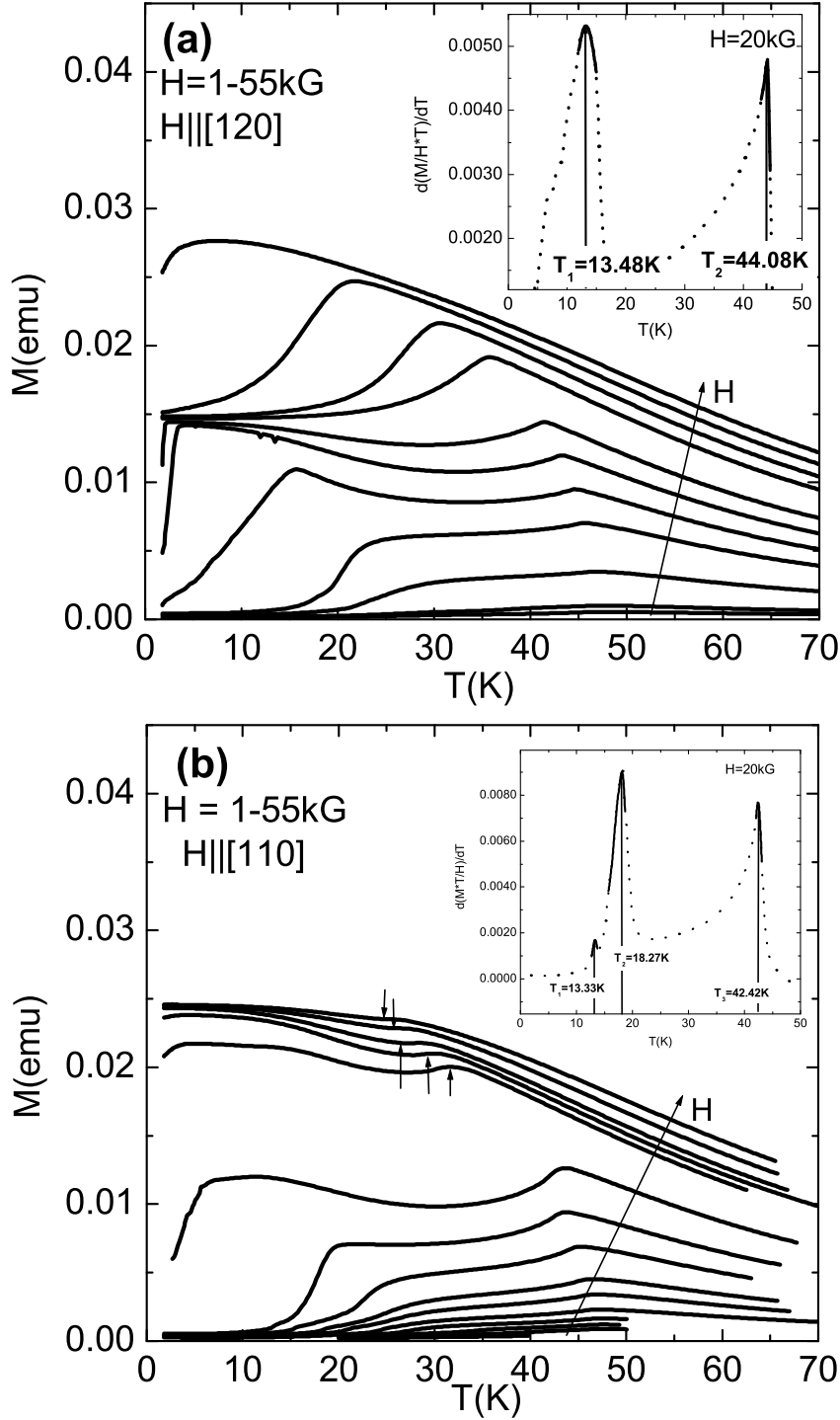}

\caption{ M(T) data for various fields for two in-plane
orientations of the applied field: (a) H$\parallel$[120] for H =
1, 2, 7.5, 15, 20, 25, 30, 40, 45, 50 and 55 kG, and (b)
H$\parallel$[110] for H = 1, 2.5, 3.5, 5, 7.5, 10, 15, 20, 26, 30,
37.5, 40, 42.5, 45, 50 and 55 kG. Insets show enlarged M(T)*T
derivatives (dotted lines) for H = 20 kG, together with the
Lorentzian fits of the maxima (solid lines), to examplify how the
points represented by full symbols on the H-T phase diagrams were
determined.}
\end{center}
\end{figure}

\clearpage

\begin{figure}
\begin{center}
\includegraphics[angle=0,width=130mm]{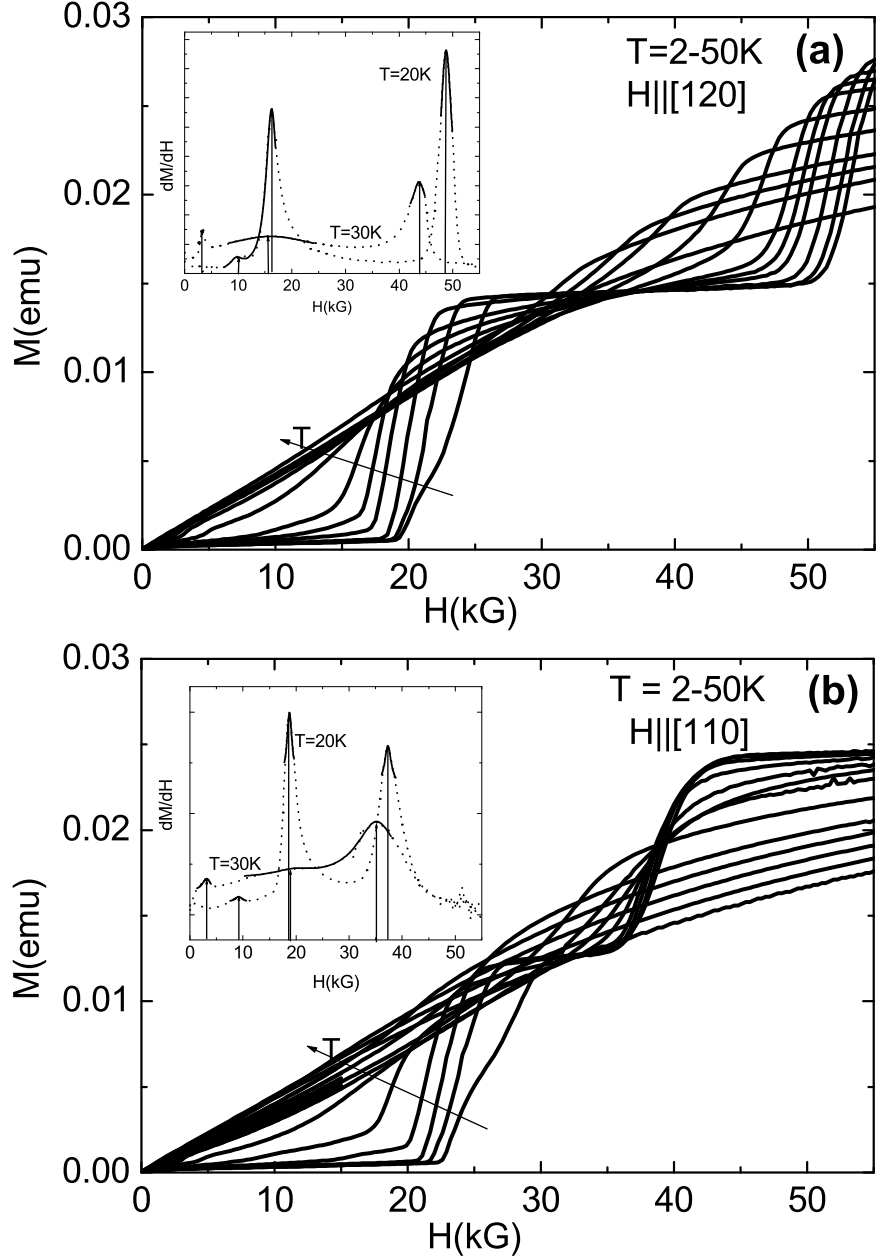}

\caption{ M(H) isotherms for two in-plane orientations of the
applied field: (a) H$\parallel$[120] for T = 3, 5, 10, 15, 17.5,
20, 25, 30, 35, 37.5, 40 and 45 K, and (b) H$\parallel$[110] for T
= 2, 5, 10, 15, 20, 25, 30, 35, 40, 42.5, 45 and 47.5 K. Insets
show enlarged M(H) derivatives (dotted lines) for T = 20 K and 30
K, for the corresponding field directions, together with the
Lorentzian fits of the maxima (solid lines), to examplify how the
points represented by open symbols on the H-T phase diagrams were
determined.}
\end{center}
\end{figure}

\clearpage

\begin{figure}
\begin{center}
\includegraphics[angle=0,width=190mm]{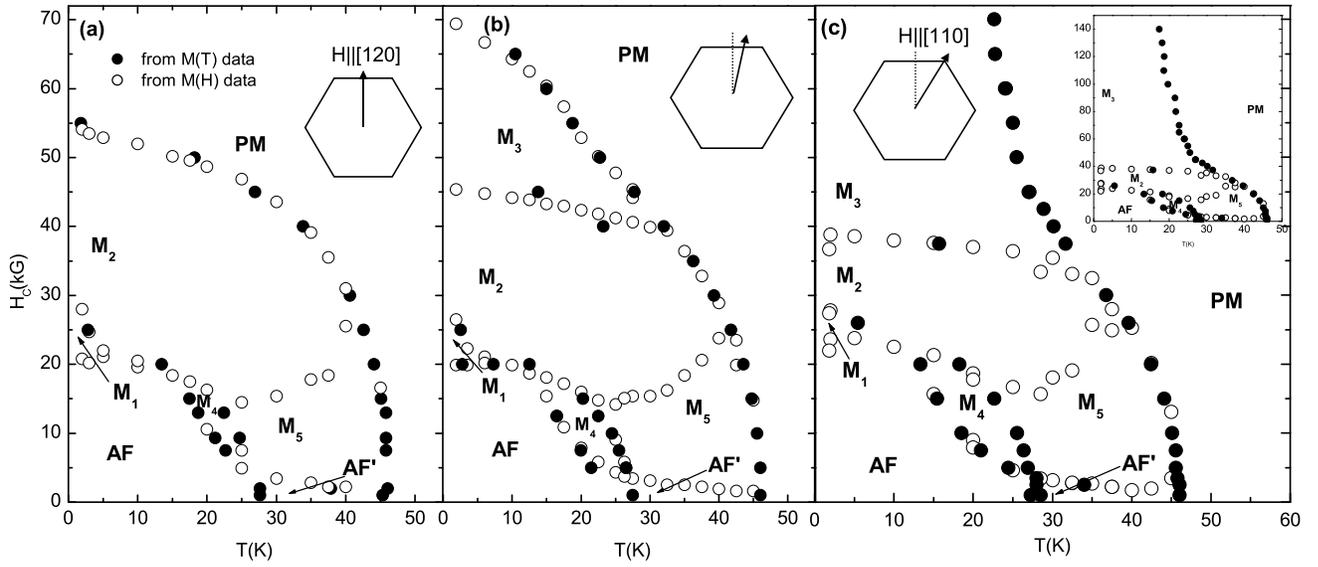}

\caption{ H-T phase diagrams for (a) H$\parallel$[120] and (c)
H$\parallel$[110], as determined from the magnetization data in
Fig.7-8, with an intermediate field orientation phase diagram (see
text) shown in (b).}
\end{center}
\end{figure}

\clearpage

\begin{figure}
\begin{center}
\includegraphics[angle=0,width=130mm]{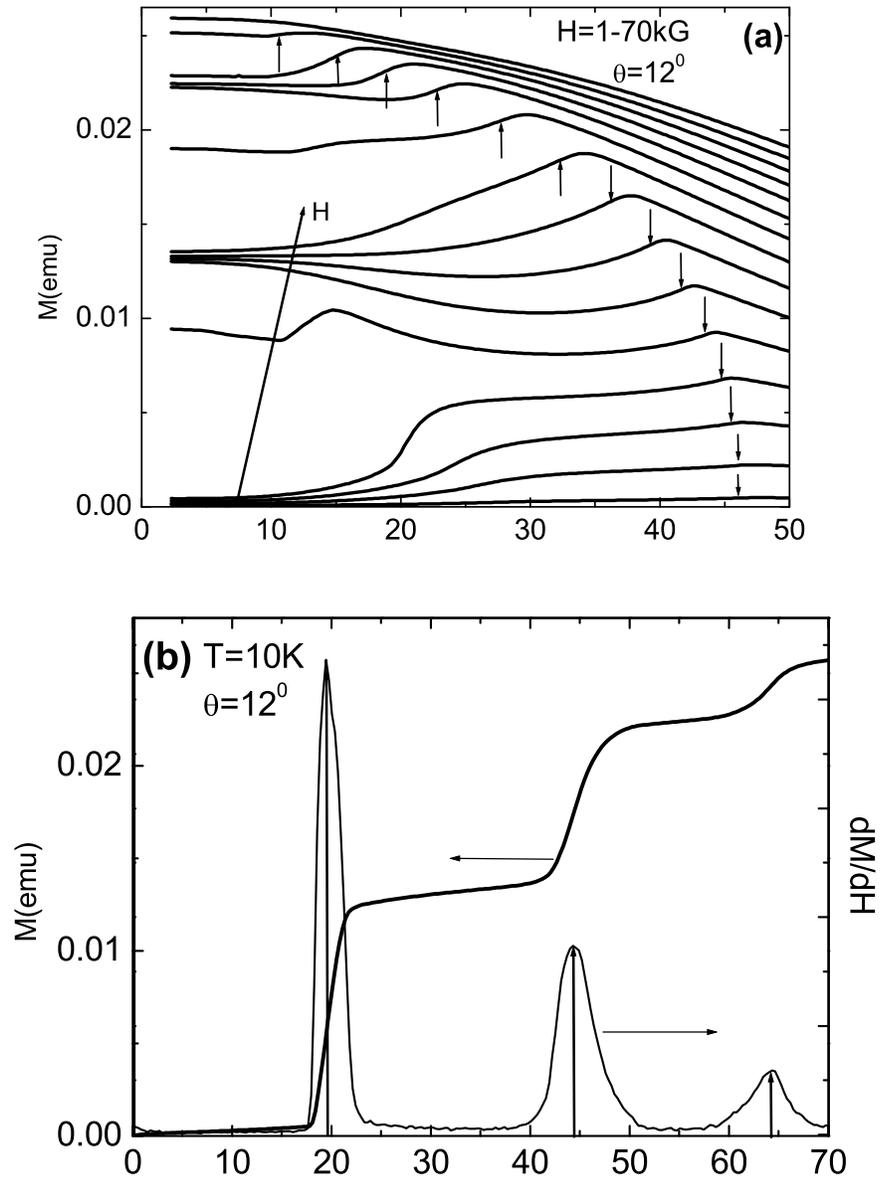}

\caption{(a) M(T) data for various fields: H = 1 kG and 5-70 kG
($\Delta$H = 5 kG) for the intermediate field direction (see
text). The small arrows indicate the highest-field transitions at
the corresponding temperature. (b) M(H) for T = 10 K, and the
corresponding derivative, as an example of how the open symbols on
the upper-most phase line in Fig.9b have been determined.}
\end{center}
\end{figure}

\clearpage

\begin{figure}
\begin{center}
\includegraphics[angle=0,width=130mm]{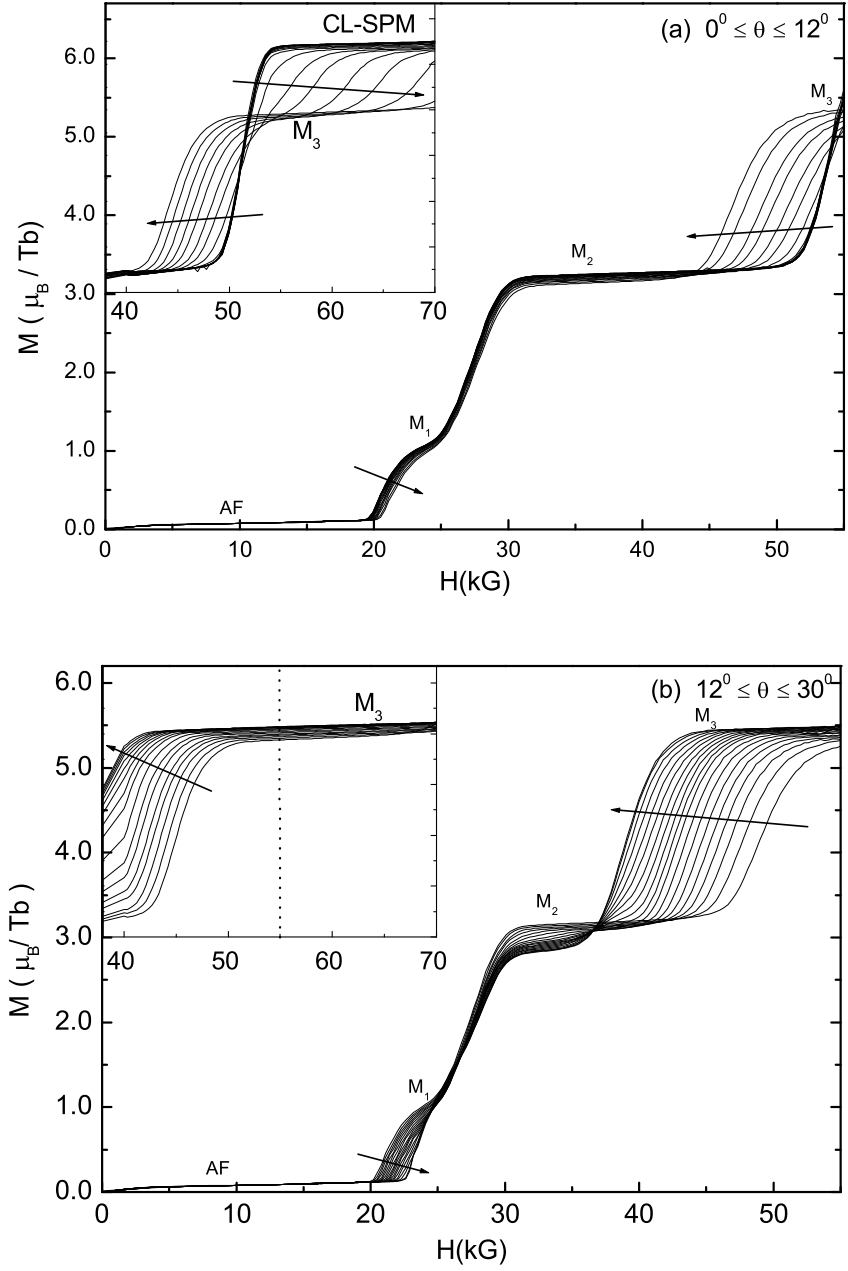}

\caption{ M(H) isotherms at T = 2.0 K for (a)
$0^0~\leq~\theta~\leq~12^0$ and (b) $12^0~\leq~\theta~\leq~30^0$
($\Delta\theta~=~1^0 $); inset: enlarged high field, T = 1.85 K
(see text) isotherms. Arrows indicate the direction of increasing
$\theta$.}
\end{center}
\end{figure}

\clearpage

\begin{figure}
\begin{center}
\includegraphics[angle=0,width=150mm]{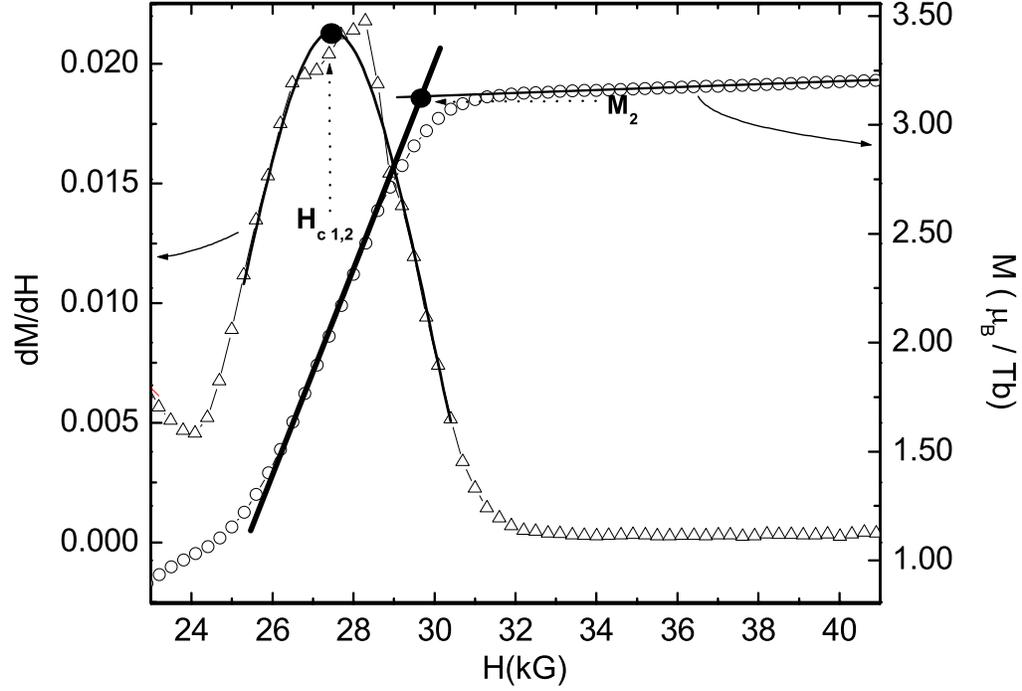}

\caption{ Enlarged M(H; $\theta~=~12^0$) plot, and the
corresponding derivative, illustrating the criteria used in
determining the points in Fig.13: large dot on the dM/dH plot
indicates the critical field value, determined by the maximum of
the Lorentzian fit (solid line) of the peak; straight lines on the
M(H) plot are fits to the magnetization plateau (thin),
extrapolated down to intersect the maximum-slope line (thick),
giving the M$_2$ value (large dot). Note: the quality of the
Lorentzian fit is representative of some of the poorer fits.}
\end{center}
\end{figure}

\clearpage

\begin{figure}
\begin{center}
\includegraphics[angle=0,width=130mm]{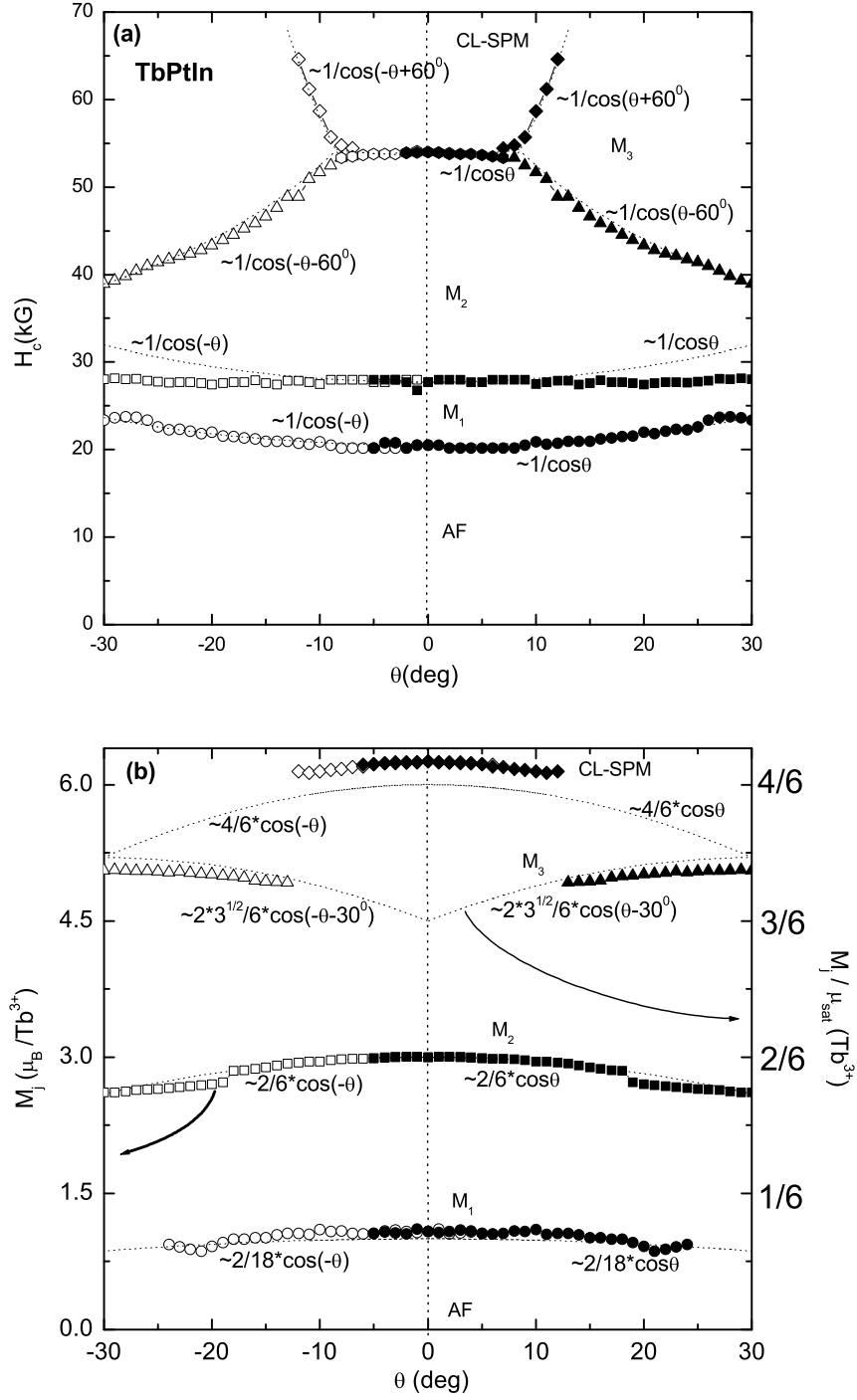}

\caption{(a) Measured critical fields H$_{ci,j}$ and (b) locally
saturated magnetizations M$_j$ (full symbols) as a function of
angle $\theta$ measured from the [120] direction. Open symbols are
reflections of the measured data across the $\theta~=~0^0$
direction. Also shown are the calculated angular dependencies
(dotted lines).}
\end{center}
\end{figure}

\clearpage

\begin{figure}
\begin{center}
\includegraphics[angle=0,width=150mm]{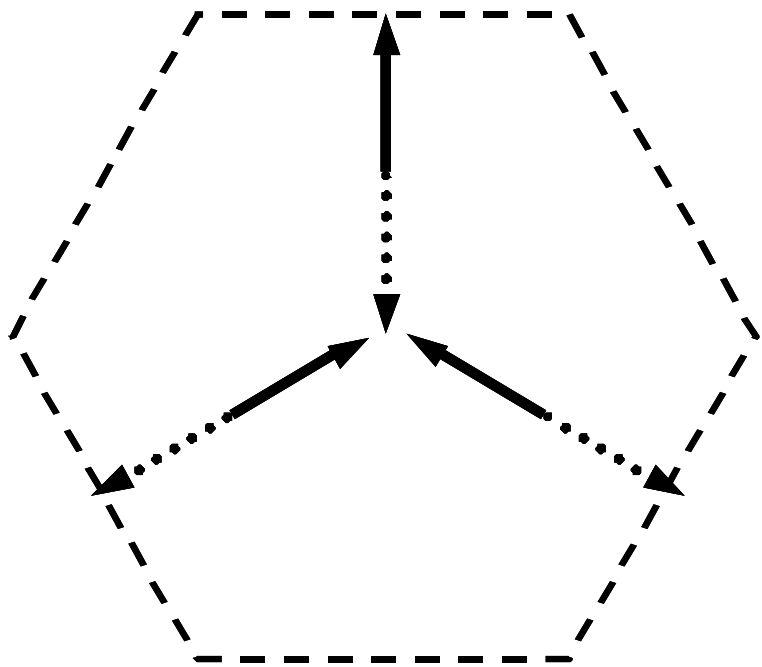}

\caption{ Schematic representation of the three Ising-like systems
model (with three distinct R in the unit cell, sitting in unique
orthorhombic sites): solid arrows- 'up', and dotted arrows- 'down'
positions of the magnetic moments along the easy axes.}
\end{center}
\end{figure}

\clearpage

\begin{figure}
\begin{center}
\includegraphics[angle=0,width=150mm]{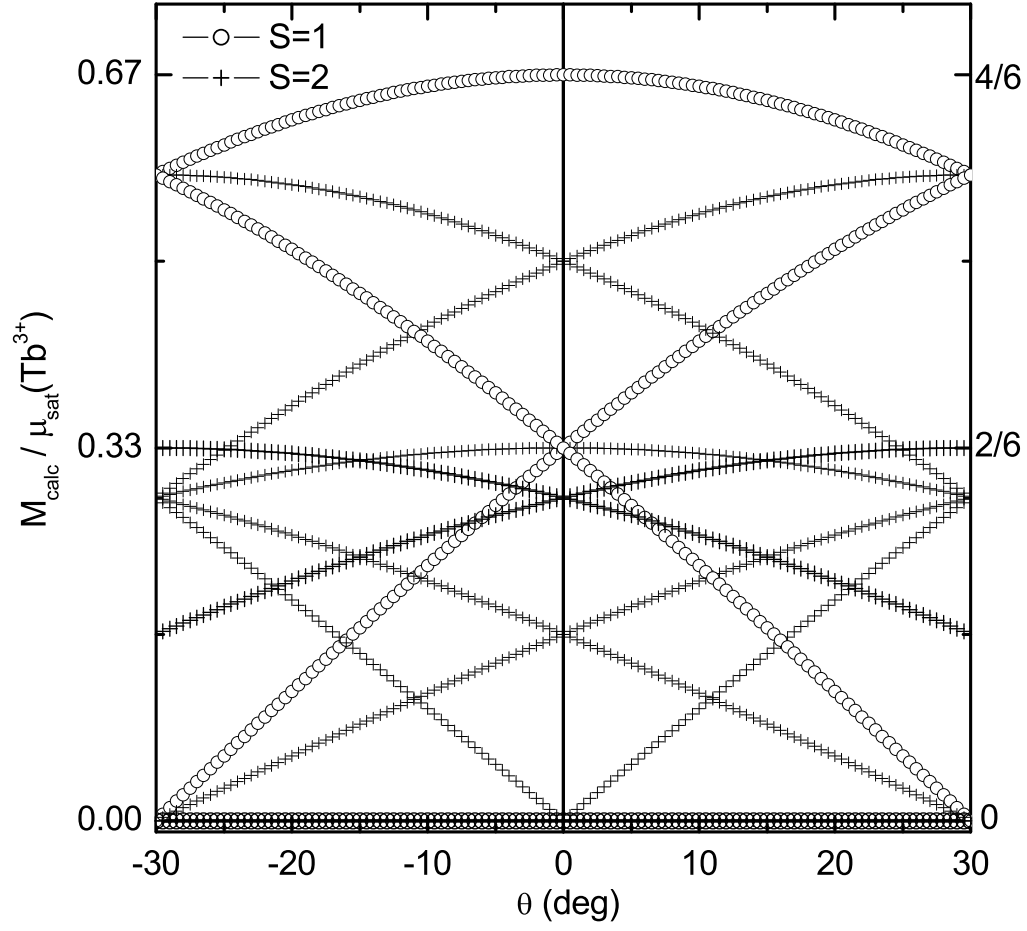}

\caption{Calculated magnetizations as a function of $\theta$,
based on the \textit{three coplanar Ising model}: open circles - S
= 1 and crosses- S = 2 (see text for details).}
\end{center}
\end{figure}

\clearpage

\begin{figure}
\begin{center}
\includegraphics[angle=0,width=150mm]{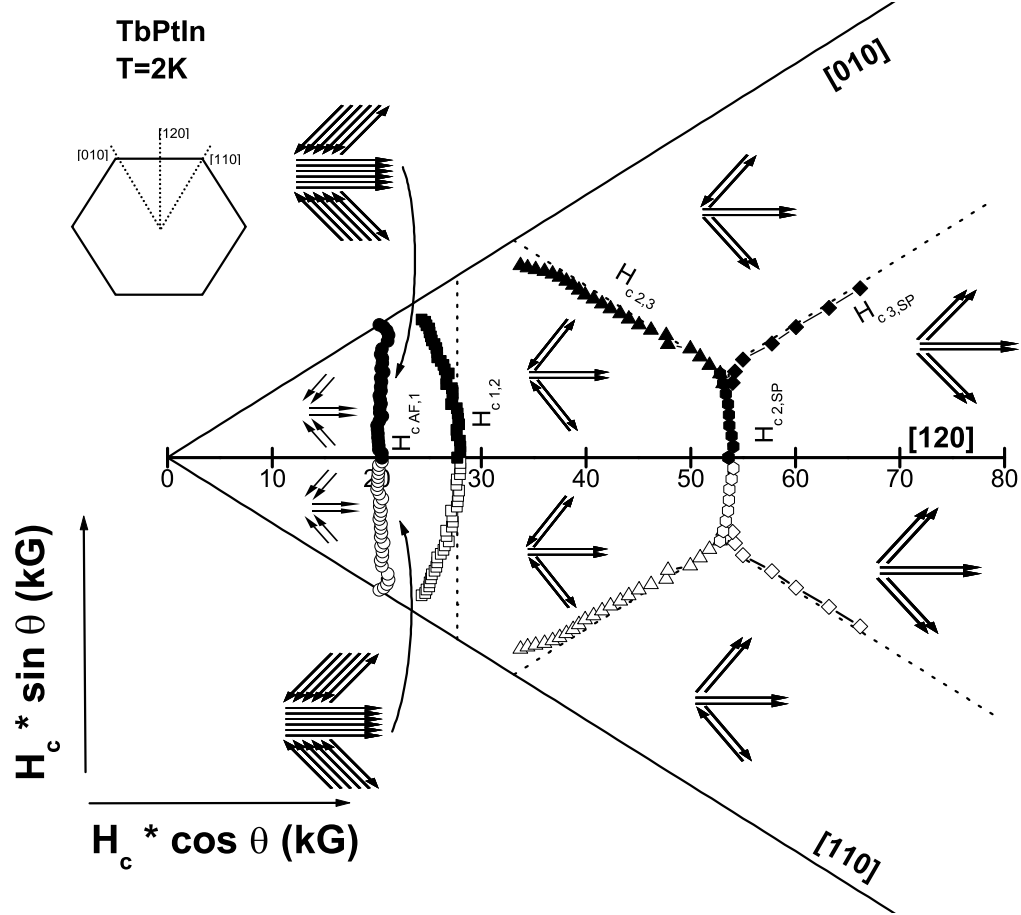}

\caption{Polar plot of the critical fields H$_{ci,j}$, with one
possible moment configuration shown for each observed metamagnetic
state; open symbols represent reflections of the measured data
-full symbols- across the $\theta~=~0^0$ direction  (see text).}
\end{center}
\end{figure}

\clearpage

\begin{figure}
\begin{center}
\includegraphics[angle=0,width=150mm]{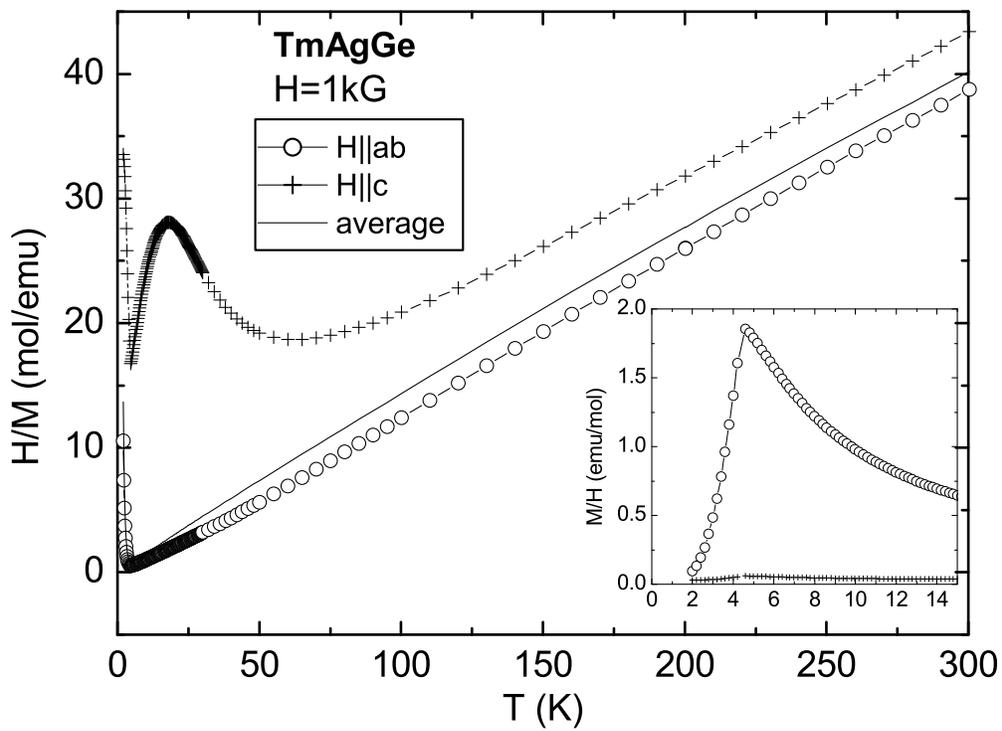}

\caption{Anisotropic inverse susceptibility of TmAgGe (symbols)
and the calculated polycrystalline average (line); inset:
low-temperature anisotropic susceptibilities.}
\end{center}
\end{figure}

\clearpage

\begin{figure}
\begin{center}
\includegraphics[angle=0,width=150mm]{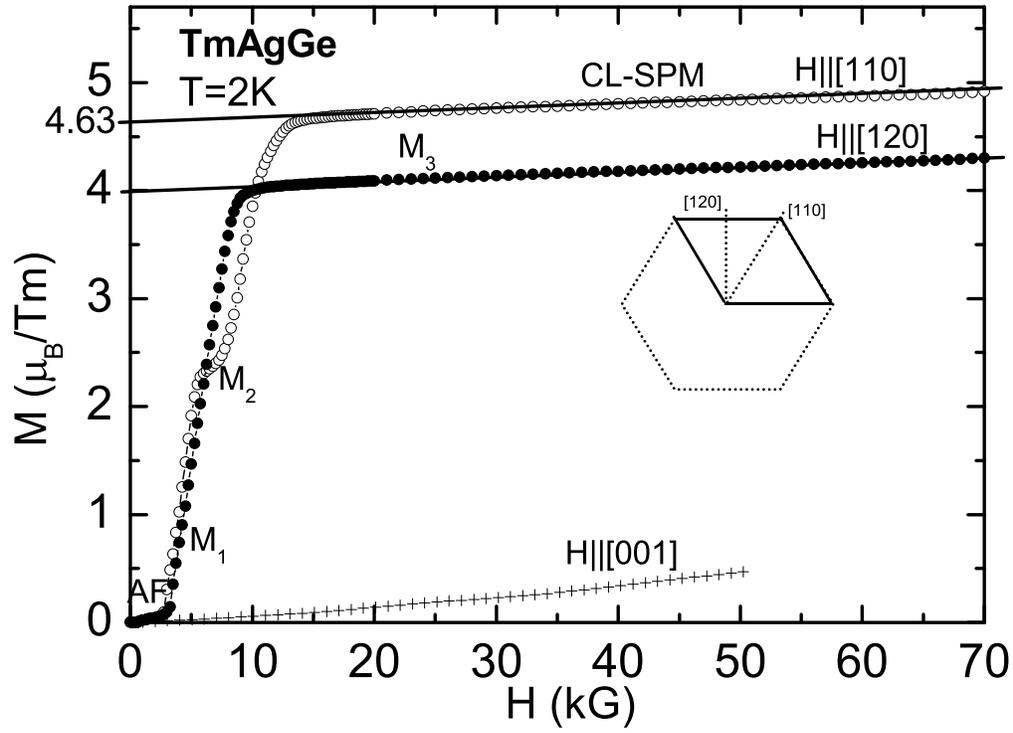}

\caption{ Anisotropic field-dependent magnetization, with the
solid lines being the linear fits of the high-field plateaus,
extrapolated down to H = 0 (see text).}
\end{center}
\end{figure}

\clearpage

\begin{figure}
\begin{center}
\includegraphics[angle=0,width=140mm]{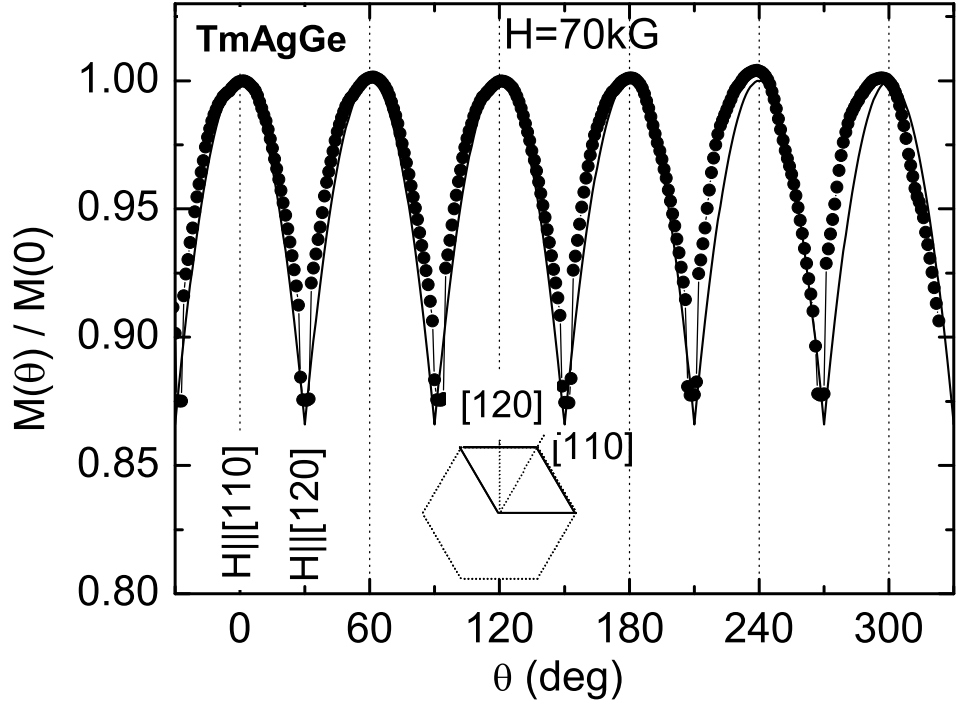}

\caption{M($\theta$) of TmAgGe (full symbols) at T = 2 K and H =
70 kG, H$\perp$c. Solid line: calculated magnetization
M$_{max}*\cos(\theta-n*60^0)$, \textit{n}-integer. Note:
$\theta~=~0^0$ is defined at the [110] direction.}
\end{center}
\end{figure}

\clearpage

\begin{figure}
\begin{center}
\includegraphics[angle=0,width=130mm]{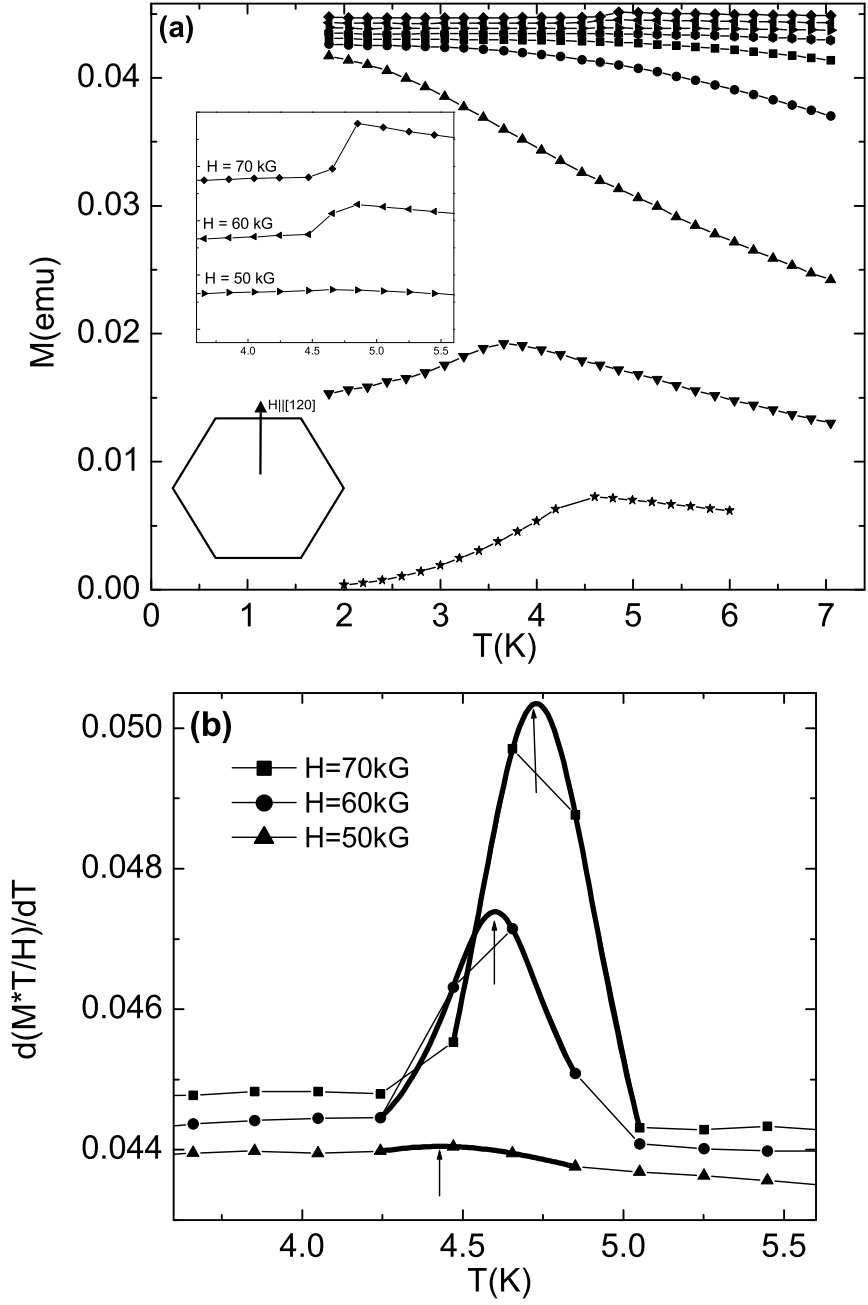}

\caption{(a) M(T) data for H = 1, 5 kG and 10 - 70 kG ($\Delta$H =
10 kG) for H$\parallel$[120], with enlarged high-field data in the
inset; (b) M*T/H derivatives for H = 50, 60 and 70 kG, together
with the Lorentzian fits of the maxima (solid lines), illustrating
how the vertical line in Fig.22c was determined for high applied
fields.}
\end{center}
\end{figure}

\clearpage

\begin{figure}
\begin{center}
\includegraphics[angle=0,width=130mm]{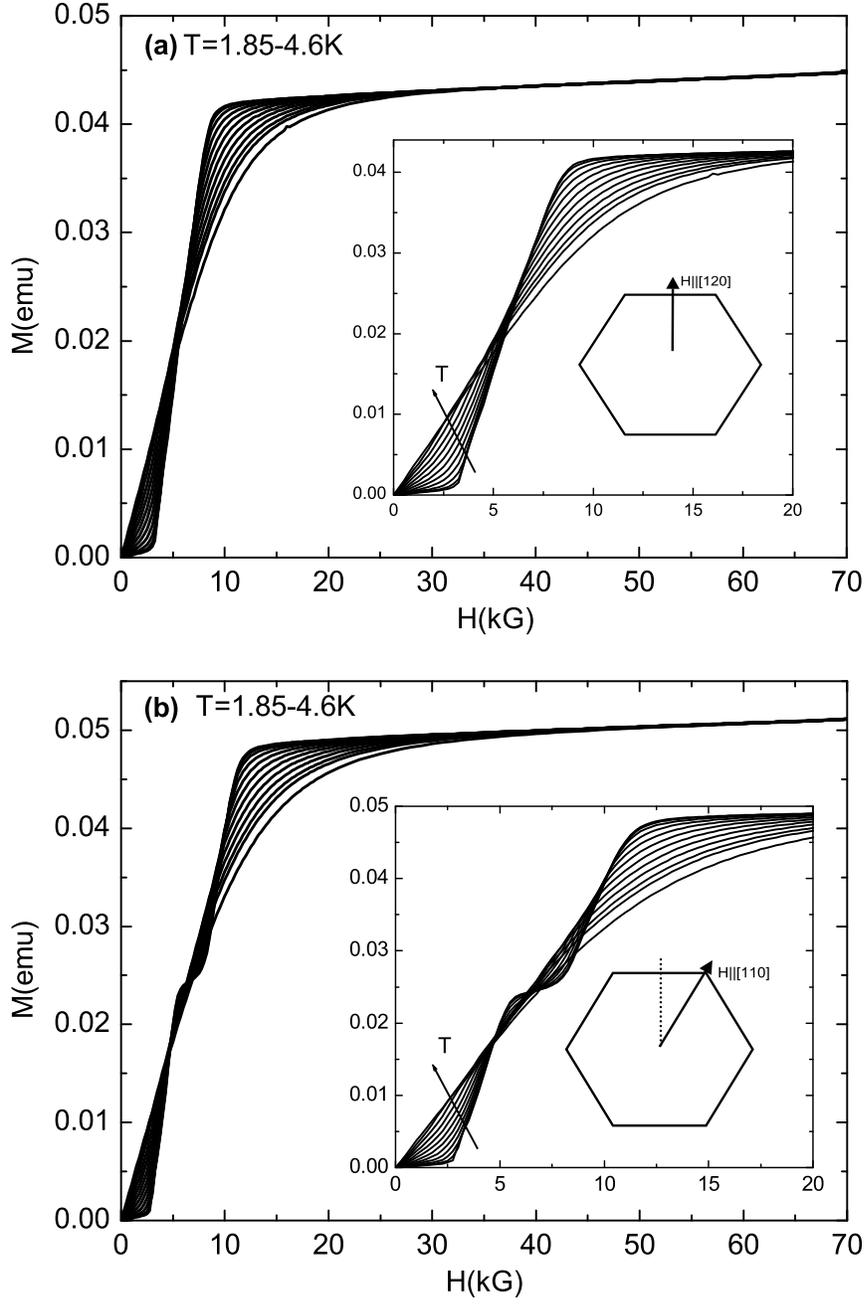}

\caption{M(H) isotherms for T $=~1.85,~2~-~4$ K ($\Delta T~=~0.25$
K), 4.2 and 4.6 K for (a) H$\parallel$[120] and (b)
H$\parallel$[110]; insets show the enlarged data around the
metamagnetic transitions. (Arrows indicate direction of increasing
T.)}
\end{center}
\end{figure}

\clearpage

\begin{figure}
\begin{center}
\includegraphics[angle=0,width=190mm]{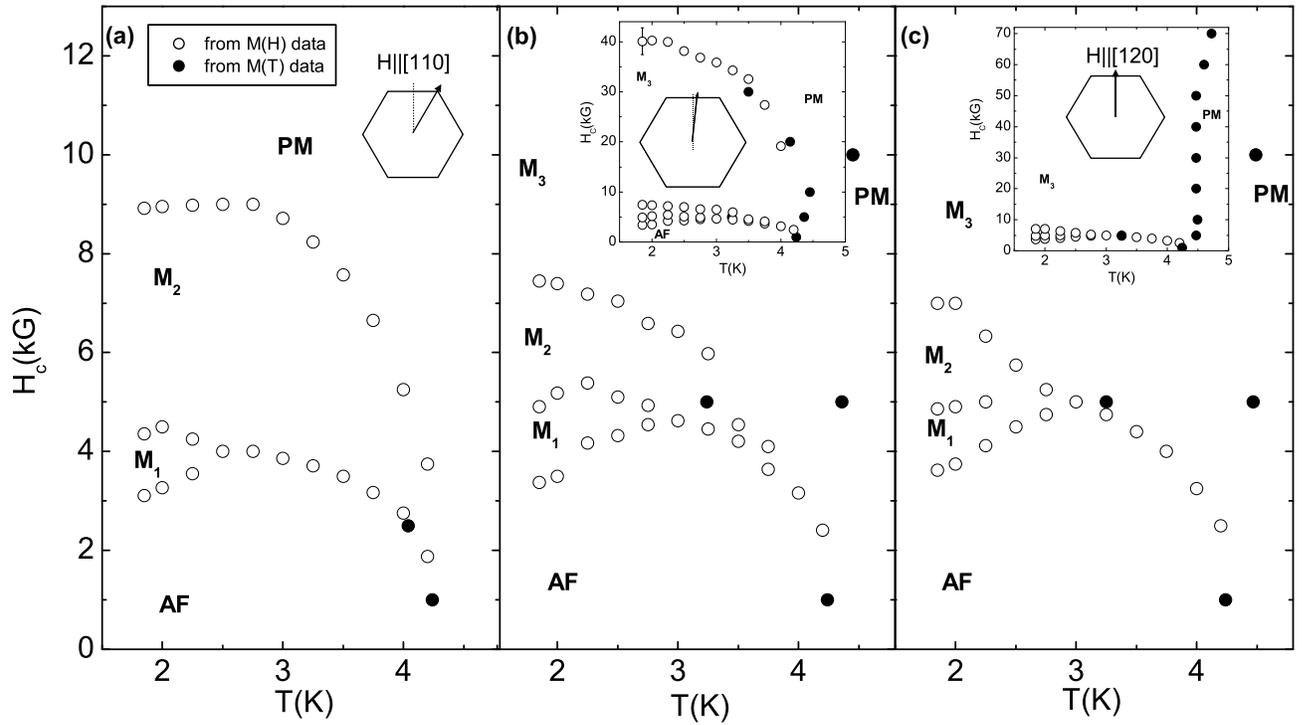}

\caption{H-T phase diagrams for TmAgGe, with (a) H$\parallel$[110]
and (c) H$\parallel$[120], as determined from magnetization data
in Fig.20-21; (b) intermediate-position phase diagram (see text);
the error bar shown in inset of (b) represents the range of values
for the high-field line, as determined from Fig.23.}
\end{center}
\end{figure}

\clearpage

\begin{figure}
\begin{center}
\includegraphics[angle=0,width=150mm]{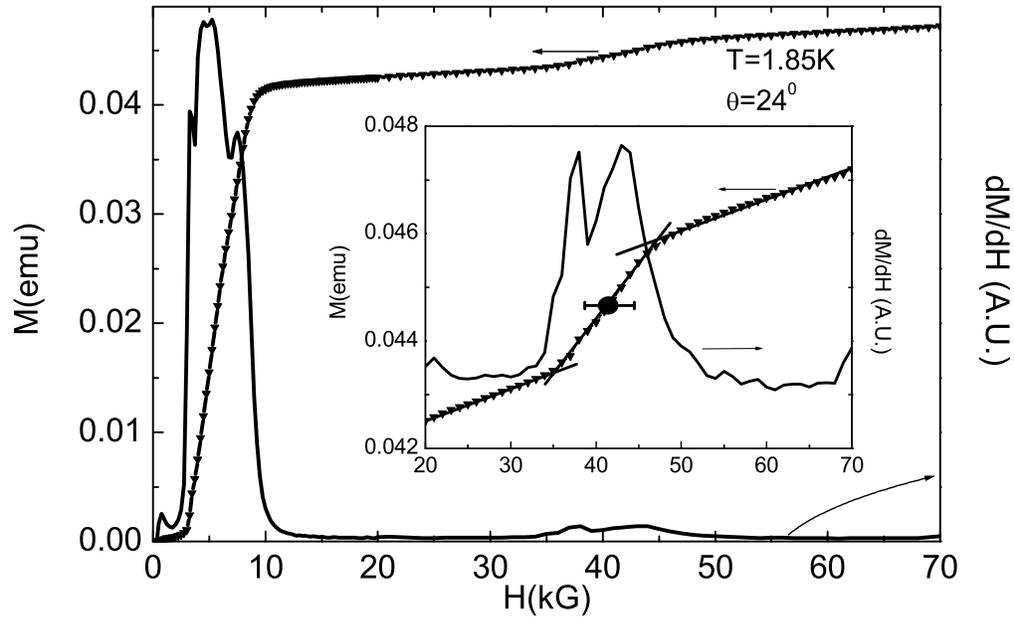}

\caption{M(H) for T = 1.85 K (symbols), and the corresponding
derivative (line). The inset shows an enlargement around the high
field transition, to exemplify the two criteria used for
determining this critical field (see text). Note: the error bar
shown in the inset (determined from the position of the peaks in
the derivative) gives a caliper of the uncertainty in determining
this critical field value.}
\end{center}
\end{figure}

\clearpage

\begin{figure}
\begin{center}
\includegraphics[angle=0,width=130mm]{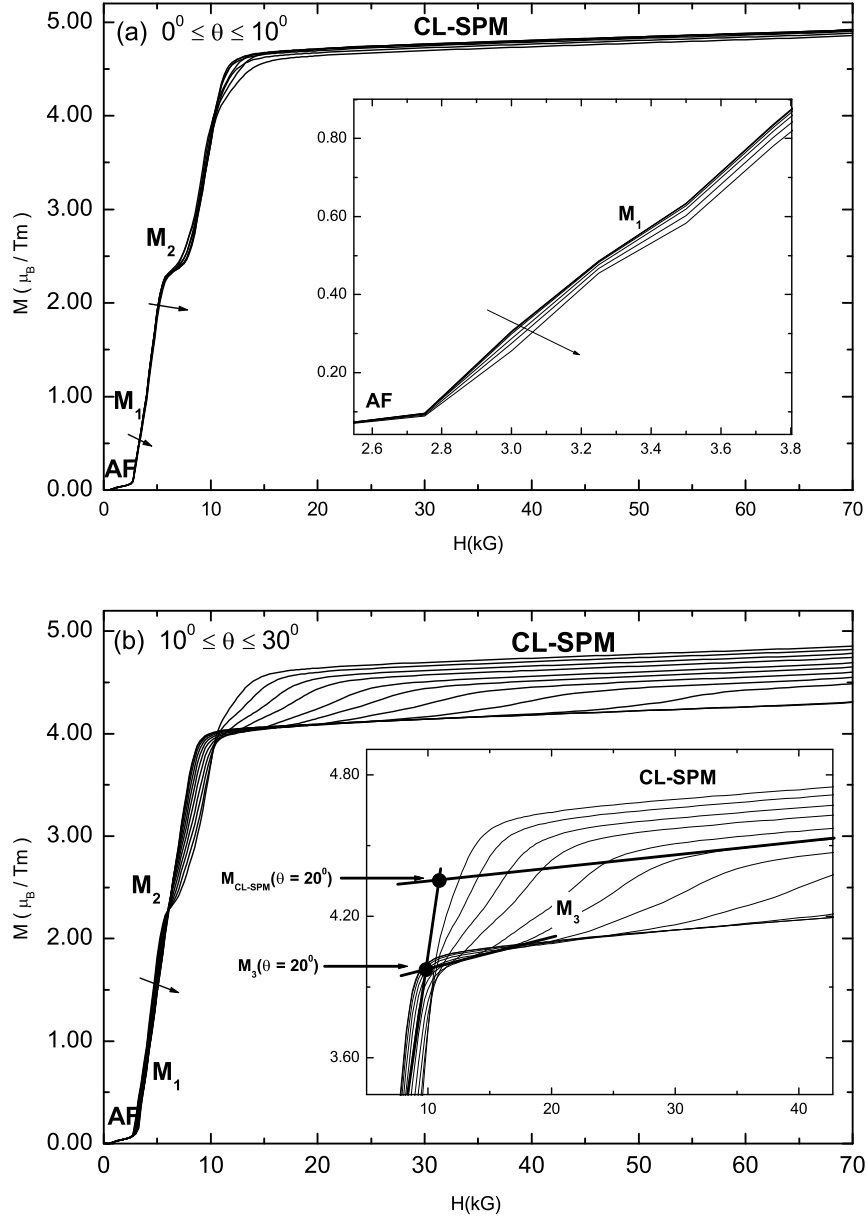}

\caption{M(H) isotherms ($T~=~2.0$ K) for (a)
$0^0~\leq~\theta~\leq~10^0$, $\Delta\theta~=~2^0$ (enlarged M$_1$
state shown in the inset) and (b) $10^0~\leq~\theta~\leq~30^0$,
$\Delta\theta~=~2^0$; inset in (b) shows the extrapolation of the
two higher metamagnetic states down to lower fields, such that the
intersection with the maximum-slope line gives the magnetization
values M$_3$ and M$_{CL-SPM}$- solid dots (see text). Arrows
indicate increasing $\theta$.}
\end{center}
\end{figure}

\clearpage

\begin{figure}
\begin{center}
\includegraphics[angle=0,width=130mm]{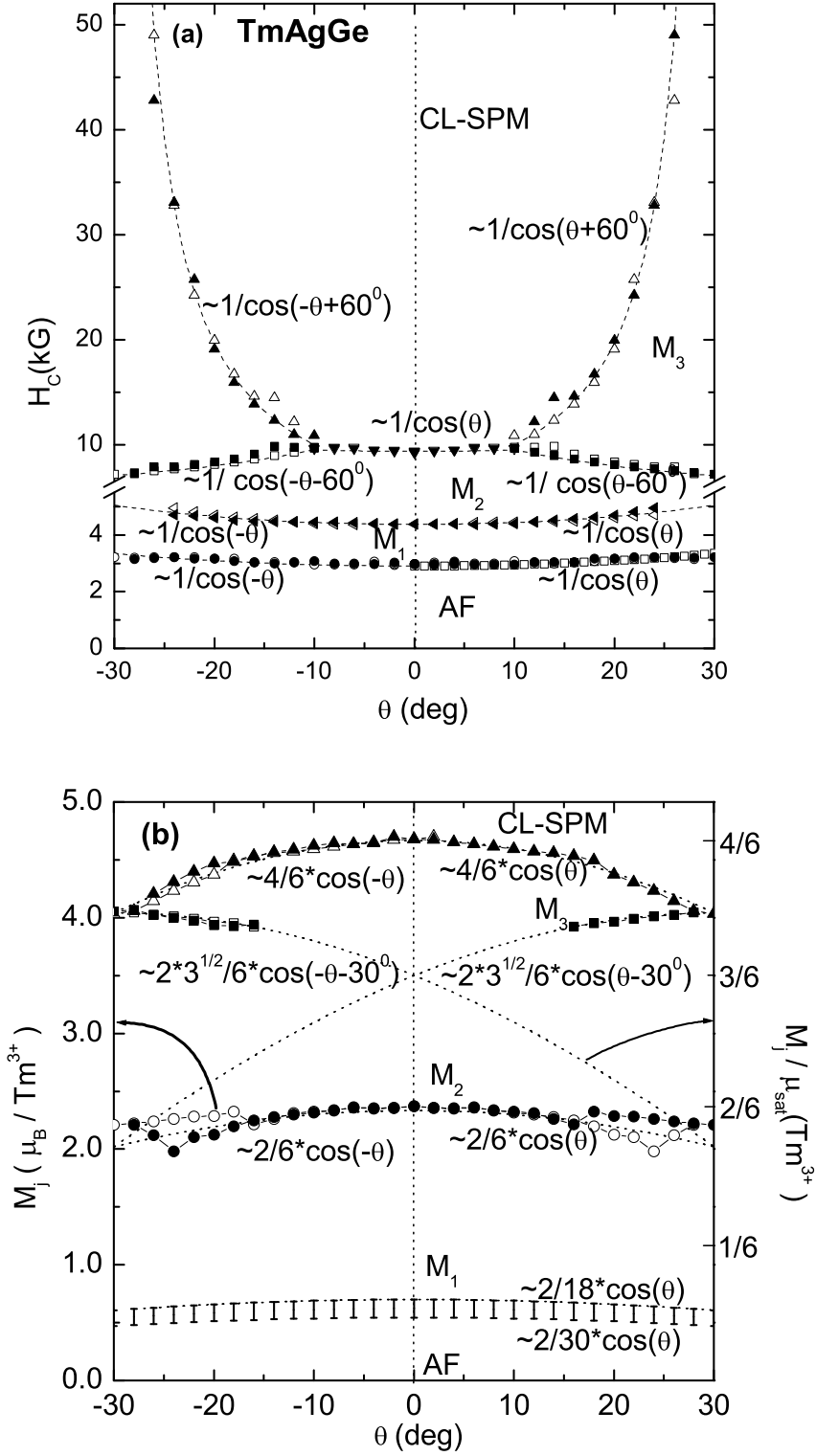}

\caption{(a) Measured critical fields $H_{ci,j}$ and (b) locally
saturated magnetizations $M_j$ (full symbols), as a function of
angle $\theta$ measured from the easy axis. Open symbols are
reflections across the $\theta~=~0^0$ direction (see text). Also
shown are the calculated angular dependencies of $H_{ci,j}$ and
$M_j$ (dotted lines). The error bars shown in the low part of (b)
give the range of values that we can infer for $M_1$ from the data
shown in Fig.24.}
\end{center}
\end{figure}

\clearpage

\begin{figure}
\begin{center}
\includegraphics[angle=0,width=130mm]{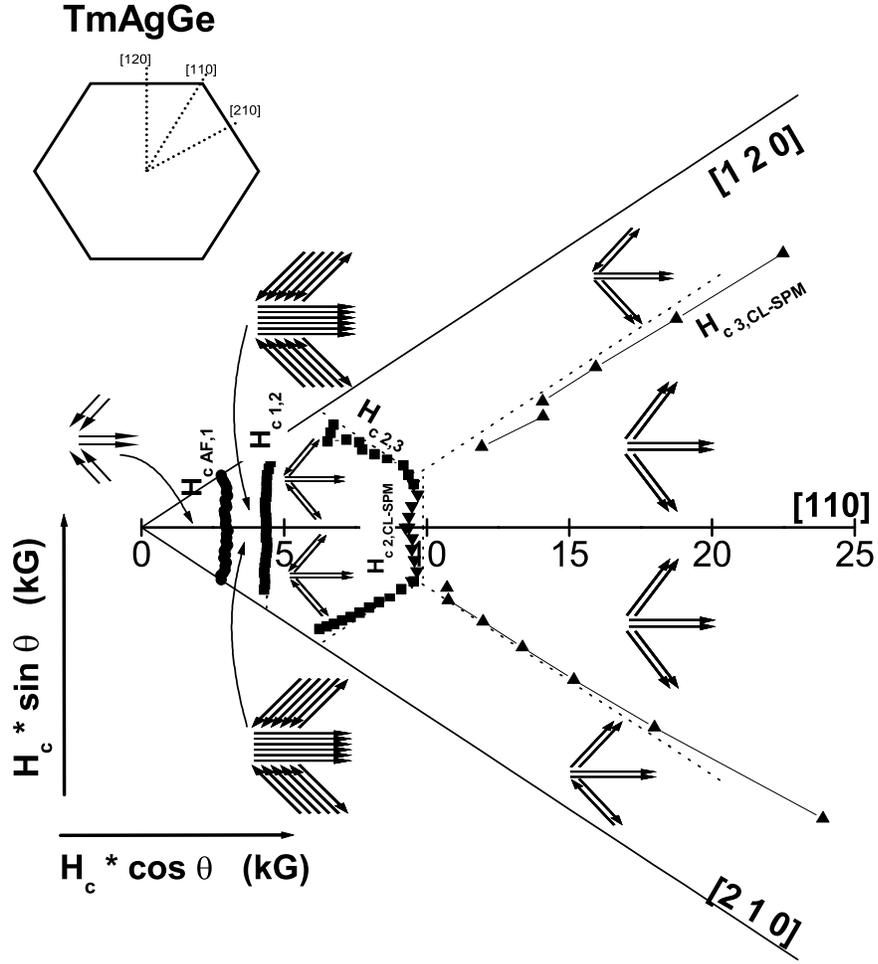}

\caption{ Polar plot of the critical fields $H_c$, with one of the
possible moment configurations shown for each observed
metamagnetic state.}
\end{center}
\end{figure}

\clearpage

\begin{figure}
\begin{center}
\includegraphics[angle=0,width=130mm]{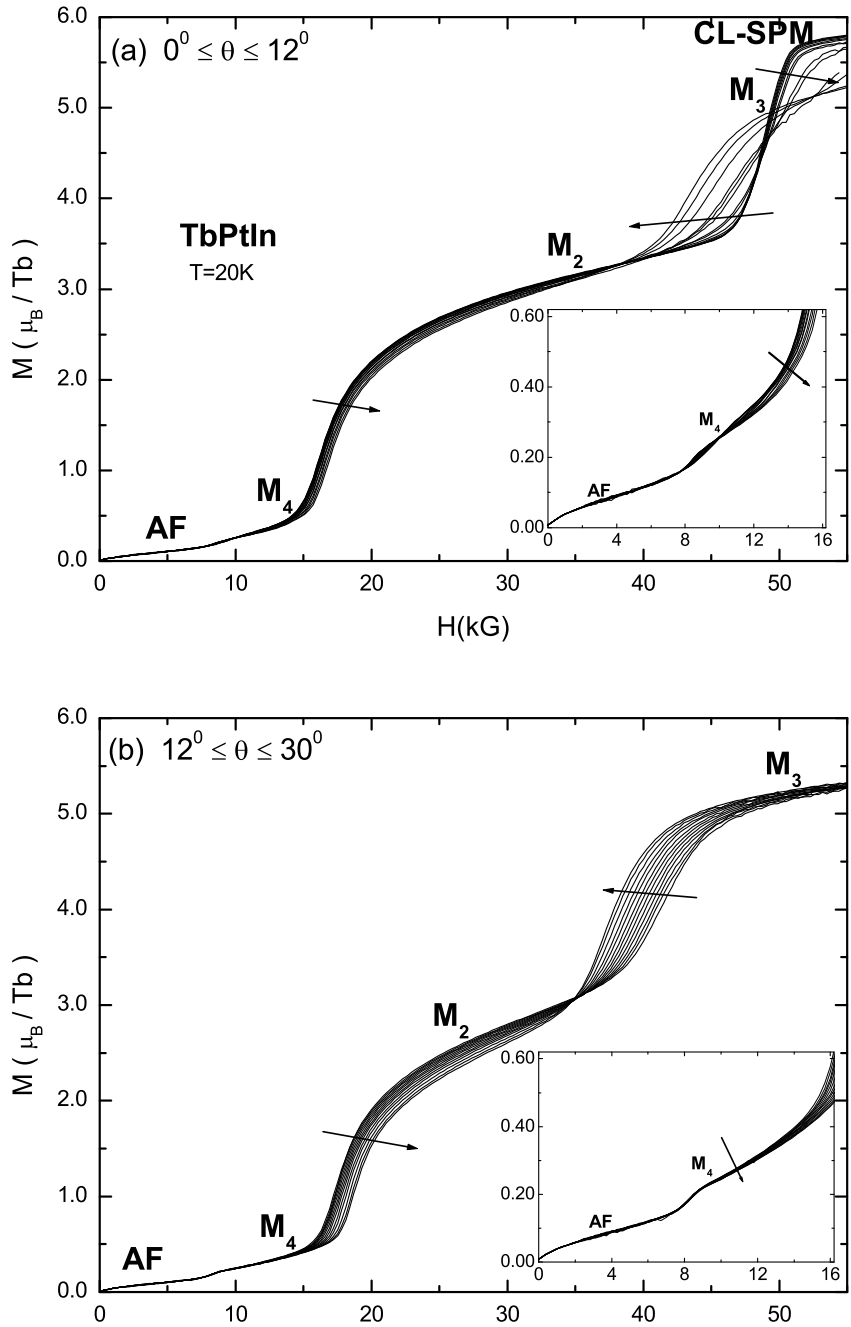}

\caption{ M(H) isotherms at $T~=~20$ K for (a)
$0^0~\leq~\theta~\leq~12^0$ and (b) $12^0~\leq~\theta~\leq~30^0$
($\Delta\theta~=~1^0 $); inset: enlarged $M_4$ state. Arrows
indicate the direction of increasing $\theta$.}
\end{center}
\end{figure}

\clearpage

\begin{figure}
\begin{center}
\includegraphics[angle=0,width=130mm]{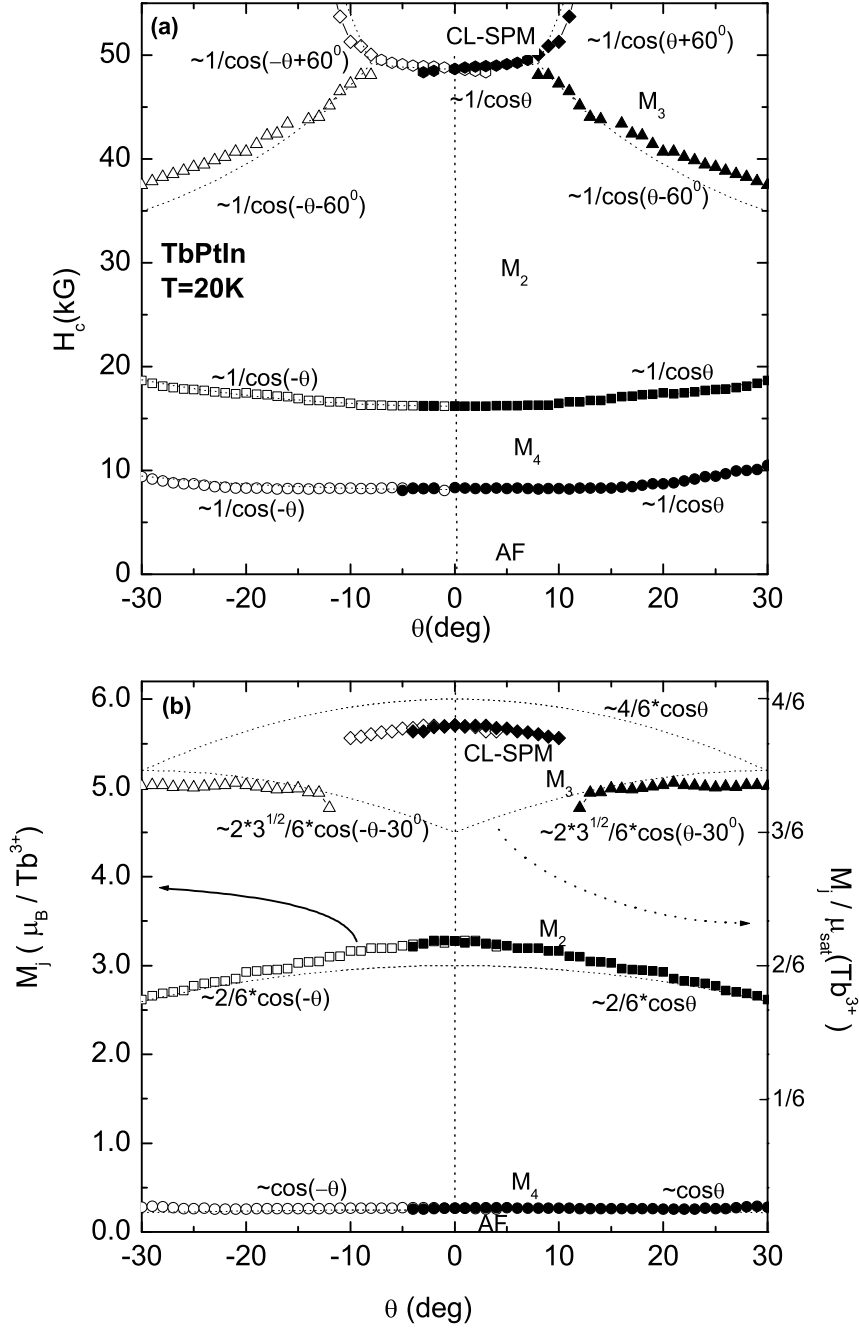}

\caption{(a) Measured critical fields $H_{ci,j}$ and (b) locally
saturated magnetizations $M_j$ (full symbols), as a function of
angle $\theta$, for $T~=~20$ K. Open symbols are reflections
across the $\theta~=~0^0$ direction. Also shown are the calculated
angular dependencies of $H_{ci,j}$ and $M_j$ (dotted lines).}
\end{center}
\end{figure}

\clearpage

\begin{figure}
\begin{center}
\includegraphics[angle=0,width=130mm]{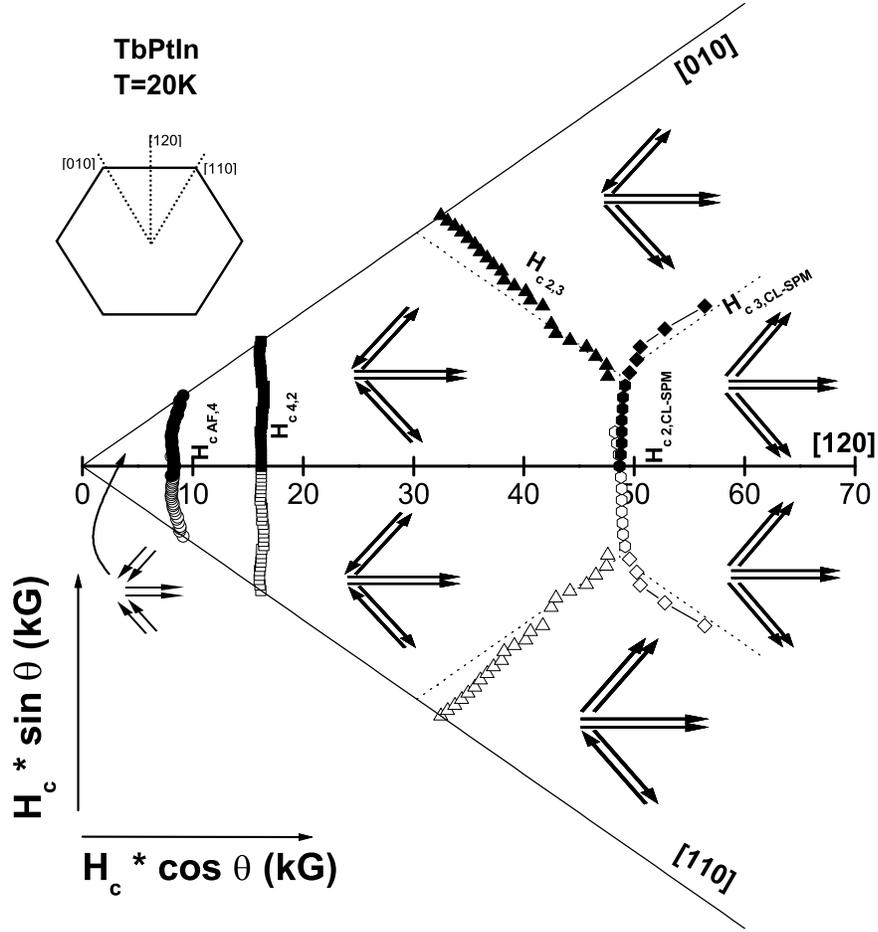}

\caption{Polar plot of the critical fields $H_{ci,j}$, with one
possible moment configuration shown for each observed metamagnetic
state (except for $M_1$, where the moment configuration is
uncertain- see text); open symbols represent reflections of the
measured data-full symbols- across the $\theta~=~0^0$ direction.}
\end{center}
\end{figure}

\clearpage

\begin{figure}
\begin{center}
\includegraphics[angle=0,width=190mm]{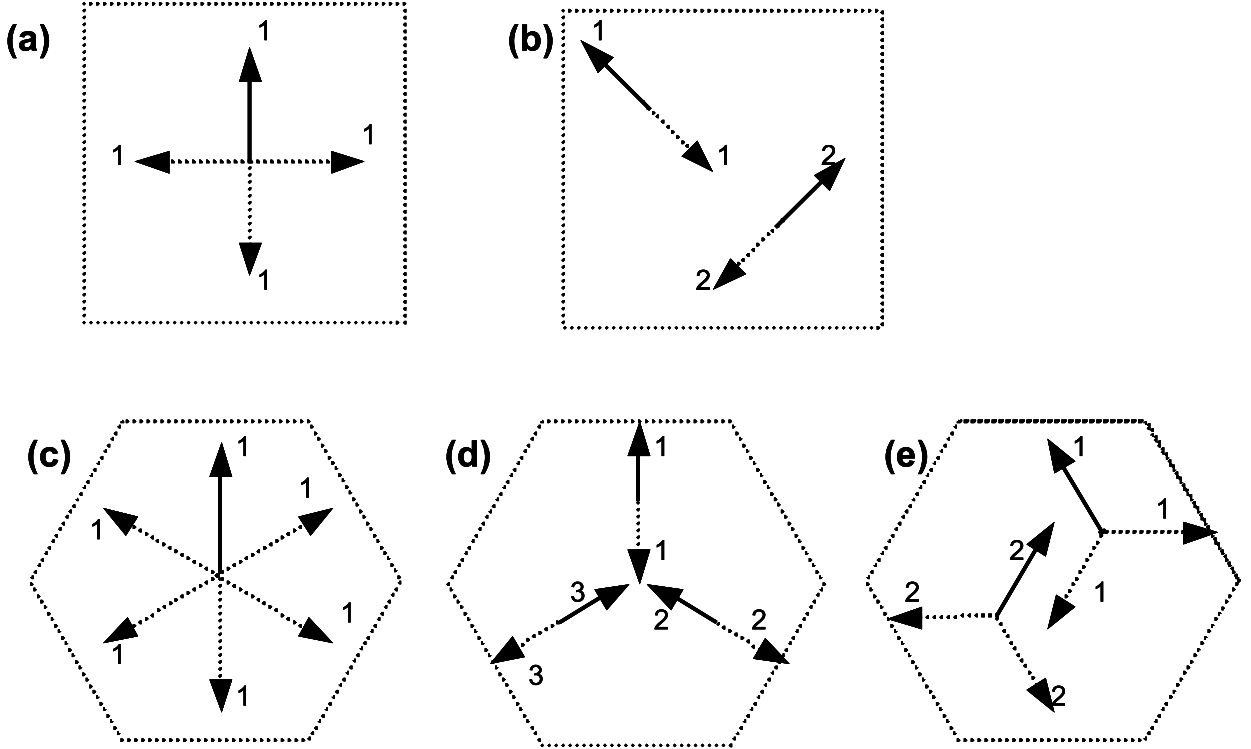}

\caption{Plausible models for extremely anisotropic, planar
compounds (see text) with (a-b) tetragonal and (c-e) hexagonal
unit cells. The numbers are used to identify the different
magnetic moments in the unit cell. (a) \textit{Four position clock
model} describing tetragonal systems with one R in tetragonal
point symmetry; (b) \textit{Double coplanar Ising-like model}: for
orthorhombic point symmetry in tetragonal unit cell, two magnetic
moments would be necessary (Ising-like systems, $90^0$ away from
each other in the basal plane). (c) \textit{Six position clock
model}: one magnetic moment with six possible orientations (arrows
along six high symmetry orientations in the basal plane); (d)
\textit{Triple coplanar Ising-like model}: three R ions in unique
orthorhombic point symmetry are needed in a hexagonal compound
(three Ising-like systems, $60^0$ away from each other in the
basal plane); (e) A \textit{double coplanar three position clock
model} can describe hexagonal systems with two magnetic moments in
unique trigonal point symmetry position, with three possible
orientations ($120^0$ away from each other in the basal plane) for
each. In all cases, the corresponding CL-SPM states (described in
text) are represented by full arrows (applied field is assumed to
be vertically up.}
\end{center}
\end{figure}

\end{document}